\documentclass[a4paper]{article}
\usepackage[a4paper,top=3cm,bottom=2cm,left=3cm,right=3cm,marginparwidth=1.75cm]{geometry}
\usepackage{amsfonts}
\usepackage{bm}
\usepackage{extarrows}
\usepackage{amssymb}
\usepackage{color}
\usepackage[all]{xy}
\usepackage{graphicx}
\usepackage{braket}
\usepackage{amsmath}
\usepackage{appendix}
\usepackage{graphicx}
\usepackage[colorinlistoftodos,textsize=tiny]{todonotes}
\usepackage{amsthm}
\usepackage{mathrsfs}
\usepackage{amssymb}
\usepackage{accents}
\usepackage{bbm}
\usepackage{extarrows}
\usepackage{makecell}
\usepackage{authblk}
\usepackage{url}
\usepackage{graphicx}
\usepackage{hyperref}
\usepackage{cite}
\usepackage{eufrak}
\hypersetup{colorlinks,
            citecolor=black,
            linkcolor=black,
            urlcolor=green,
            pdftex}

\def\be{\begin{equation}}
\def\ee{\end{equation}}
\def\ba{\begin{eqnarray}}
\def\ea{\end{eqnarray}}

\usepackage[english]{babel}
\usepackage[utf8x]{inputenc}
\usepackage[T1]{fontenc}

\usepackage[normalem]{ulem}

\title{Holonomy operator for spin connection and spatial scalar curvature operator in loop quantum gravity}
\author[1,2]{Gaoping Long \footnote{201731140005@mail.bnu.edu.cn}}
\author[3]{Hongguang Liu \footnote{hongguang.liu@gravity.fau.de}\thanks{corresponding author}}

\affil[1]{College of Physics $\&$ Optoelectronic Engineering, Jinan University, Guangzhou, 510632, Guangdong, China}
\affil[2]{Department of Physics, Beijing Normal University, Beijing, 100875, China}

\affil[3]{Department Physik, Institut f\"ur Quantengravitation, Theoretische Physik III, Friedrich-Alexander Universit\"at Erlangen-N\"urnberg, Staudtstr. 7/B2, 91058 Erlangen, Germany}

\date{}

\begin{document}

\maketitle

\begin{abstract}
In this article we propose a new construction of the spatial scalar curvature operator in (1+3)-dimensional LQG based on the twisted geometry. The starting point of the construction is to express the holonomy of the spin connection on a graph in terms of the twisted geometry variables, and we check that this expression reproduces the spin connection in terms of triads in a certain continuum limit. The spatial scalar curvature in terms of twisted geometry is obtained by considering the composition of the holonomy of the spin connection on the loops. With the twisted geometry parametrization of the holonomy-flux phase space, we further express the holonomy of the spin connection and the spatial scalar curvature on a graph in terms of fluxes. Finally, they are promoted as well-defined operators by replacing the fluxes with ordered flux operators.
\end{abstract}

\section{Introduction}

\label{sec:Intro}
Loop quantum gravity (LQG) is a promising approach to non-perturbative and background independent quantum gravity  \cite{thiemann2008modern,Ashtekar:2004eh,rovelli_vidotto_2014,Han2005FUNDAMENTAL}. This theory starts from the Hamiltonian formulation of (1+3)-dimensional general relativity (GR), which is formulated as a Yang-Mills gauge theory, with the Ashtekar-Barbero connection and desitized triads on 3-dimensional spatial slice being the conjugate pairs and Gaussian constraint generating the gauge transformations. The quantum states of this theory are spin-network states, which provide the basic building blocks of the discrete quantum geometry. Also, this quantization framework is extended to all dimensional GR, and it leads to similar kinematical structures which describe the discrete quantum geometry in arbitrary spacetime \cite{Bodendorfer:Ha,Bodendorfer:Qu,long2019coherent,long2020operators}.
Based on classical deparametrized models of gravity, such as gravity coupled to dust fields or scalar fields, the reduced phase space quantization is proposed for LQG \cite{Giesel:2007wn,Thiemann:2023zjd,Brown:1994py,PhysRevD.43.419,Domagala:2010bm}. With the dynamical constraints being solved classically and the coupled dust fields or scalar fields being used to parametrize the spacetime coordinates, the reduced phase space quantization has a true physical Hamiltonian instead of a Hamiltonian constraint, and the gauge invariant Hilbert space with respect to the Gaussian constraint in LQG becomes the physical Hilbert space of Dirac observables.
Correspondingly, the quantum dynamics is governed by the entire LQG physical Hamiltonian defined on the physical Hilbert space. One of the popular Hamiltonians is given by the Giesel-Thiemann construction, where the Hamiltonian constraint is separated into Euclidean and Lorentzian terms.
Thiemann's method is employed to quantize the Lorentzian term in relation to the extrinsic curvature operators \cite{thiemann2008modern}. It has been shown that the semiclassical limit of the theory based on the reduced phase space coherent state path integral formulation reproduces classical reduced phase space equation of motions (EOMs) of gravity \cite{Han:2019vpw,Han:2020chr,Long:2021izw}. Thus, it is semiclassically consistent.

There is another proposal for the Hamiltonian by Alesci-Assanioussi-Lewandowski-Makinen (AALM) \cite{Alesci:2015wla,Assanioussi:2015gka}, which replaces the relatively complicated form of the Lorentzian term with the spatial scalar curvature on the 3-dimensional spatial slice. The spatial scalar curvature itself is a geometric observable characterizing the geometry of the spatial manifold. Additionally, similar formulations are widely used in symmetry-reduced models inspired by LQG, such as standard loop quantum cosmology (LQC) and loop quantum black hole models, leading to singularity resolution and big bounce \cite{Bojowald:2001xe, Ashtekar:2006rx, Ashtekar:2006wn,Ashtekar:2005qt,Modesto:2005zm, Boehmer:2007ket,Chiou:2012pg,Gambini:2013hna,Corichi:2015xia,Dadhich:2015ora,Olmedo:2017lvt,Ashtekar:2018lag,BenAchour:2018khr,Han:2020uhb,Kelly:2020lec, Han:2022rsx,Giesel:2023hys,Ashtekar:2023cod,Zhang:2021zfp,PhysRevLett.102.051301}.
Consequently, a scalar curvature operator will play an important role in deriving symmetry-reduced models from the first principle of full LQG. Moreover, this proposal is also considered in all dimensional LQG to avoid the problem that the  Euclidean term of Hamiltonian constraint cannot capture the degrees of freedom of the intrinsic curvature \cite{Long:2022thb}.

The proposed operator representing the spatial scalar curvature is quantized with Regge calculus techniques on a cellular decomposition of the spatial manifold \cite{regge,Alesci:2014aza}. However, contrary to the Giesel-Thiemann construction, the continuum limit of such an operator is not straightforward and it is not compatible with the Euclidean part of the Hamiltonian constraint on a fixed graph. The conflict is related to the fact that the Regge calculus relies on a cellular decomposition dual to the graph, and to the fact that the holonomy-flux variables on a fixed graph correspond to twisted geometry instead of Regge geometry \cite{Rovelli:2010km,Freidel:2010bw,PhysRevD.82.084040,PhysRevD.103.086016}. To avoid such a problem, a new quantization of the scalar curvature operator on a cubic graph with regularization of the densitized triad and its covariant derivatives is proposed \cite{Lewandowski:2021iun,Lewandowski:2022xox}. However, the operator is only defined on a cubical graph with a relatively complicated formula and has no intuitive discrete geometric interpretation.

The issues encountered in previous works inspire us to establish the spatial scalar curvature operator based on the twisted geometry instead of the Regge geometry.
More explicitly, the twisted geometry provides a parametrization of the holonomy-flux phase space, hence the discrete geometry captured by the holonomy-flux variables on a fixed graph is interpreted by the twisted geometry instead of the Regge geometry \cite{Rovelli:2010km,Freidel:2010bw,PhysRevD.103.086016,Long:2023ivt}. In fact, the space of Regge geometry, which is given by imposing the shape-matching condition of the 2-face in the twisted geometry space, is a subspace of the twisted geometry space on the graph dual to a cellular decomposition. Moreover, the Hilbert space spanned by spin network states on the graph is given by quantizing the corresponding holonomy-flux phase space. Therefore, it is reasonable to construct the spatial scalar curvature operator based on the twisted geometry instead of the Regge geometry. This ensures that the corresponding spatial scalar curvature operator acting on the spin network states could represent the correct quantum degrees of freedom of the discrete geometry.  A prelude for the study in this direction is given in Ref.\cite{PhysRevD.87.024038}, in which the distributional spin connection  of the discrete frame on tetrahedron  is constructed in the framework of twisted geometry. Nevertheless, this construction relies on the triangulation of the spatial manifold, and it is still an open issue to construct the spin connection in twisted geometry for generic cellular decomposition. Besides, it is also desired to construct the operators for the spin connection  and the scalar curvature based on the twisted geometry.

In this paper, we propose a new scalar curvature operator based on the holonomy operators of spin connections. First, we establish the holonomy of the spin connection in terms of the twisted geometry variables for generic cellular decomposition. Correspondingly, the spatial discrete scalar curvature on a closed graph can be established based on the holonomy of spin connections. We also check that the holonomy of the spin connection and the corresponding spatial discrete scalar curvature on the cubic graph reproduce the continuum spin connection and the spatial curvature in the continuum limit in a certain coordinate system.
Then, note the twisted geometric parameterization of the holonomy-flux phase space, the holonomy of the spin connection and the spatial discrete scalar curvature can be expressed by the fluxes. Finally, up to the operator ordering, they can be promoted as operators by simply replacing the classical fluxes by flux operators.

This paper is organized as follows. In the following section \ref{sec2} we will review the kinematical structure and the existing treatment of the Hamiltonian constraint in LQG. In particular, this section also serves to introduce the twisted geometry parametrization of the holonomy-flux phase space on a fixed graph. In section \ref{sec3} we introduce the construction of the holonomy of the spin connection and the spatial scalar curvature on a graph in terms of the twisted geometry variables, and check that these expressions reproduce correct continuum limits for the cubic graph. Then, in section \ref{sec4}, we propose the quantization of these expressions. Finally, in section \ref{sec5} we summarize and discuss the results with an outlook on the possible next steps of future research.

\section{Elements of LQG}\label{sec2}
\subsection{The basic structures}
The (1+3)-dimensional Lorentzian LQG is constructed by canonically quantizing GR based on the Yang-Mills phase space with the non-vanishing Poisson bracket\cite{Ashtekar:2004eh,Han2005FUNDAMENTAL}
 \begin{equation}
 \{A_{a}^i(x),E^{b}_j(y)\}=\kappa\beta\delta_a^b\delta^i_j\delta^{(3)}(x-y),
 \end{equation}
where the configuration and momentum variables are the  $su(2)$-valued connection field $A_{a}^i$ and the densitized triad field $E^{b}_j$ respectively on a 3-dimensional spatial manifold $\Sigma$, $\kappa=8 \pi G$ with the gravitational constant $G$,
and $\beta$ represent the Babero-Immirze parameter. Here we use $i, j, k, ...$ for the internal $su(2)$ index and $a, b, c, ...$ for the spatial index. Let $q_{ab}=e_a^ie_{bi}$ be the spatial metric on $\Sigma$. The densitized triad is related to the triad $e_i^a$ by $E^{a}_{i}=\sqrt{\det(q)}e^{a}_{i}$, where $\det(q)$ denotes the determinant of $q_{ab}$. The connection can be expressed as $A_{a}^{i}=\Gamma_{a}^{i}+\beta K_{a}^{i}$, where $\Gamma_{a}^{i}$ is the Levi-Civita connection of $e_{a}^{i}$, which is given by \cite{thiemann2008modern}
\begin{equation}
\Gamma_{a}^{i}=\frac{1}{2}\epsilon^{ijk}e_k^b
(\partial_be_a^j-\partial_ae_b^j+e^c_je_{al}\partial_be^l_c).
\end{equation}
 $K_a^i$ is related to the extrinsic curvature $K_{ab}$ by $K_a^i=K_{ab}e^b_j\delta^{ji}$. The Gaussian constraint which generates gauge transformation is given by
\begin{equation}\label{GC}
 \mathcal{G}:=\partial_aE^{ai}+A_{aj}E^a_k\epsilon^{ijk}\approx0.
\end{equation}
Again, the dynamics of LQG can be defined by the physical Hamiltonian which is introduced by the classical deparametrization of GR. In the deparametrization models with certain dust fields, the diffeomorphism and Hamiltonian constraints are solved classically and the dust reference frame provides the physical space-time coordinates, so that the theory is described in terms of Dirac observables \cite{Brown:1994py}\cite{PhysRevD.43.419}\cite{Domagala:2010bm}. Then, the physical time evolution is generated by the physical Hamiltonian with respect to the
dust fields. More explicitly, the resulting physical Hamiltonian $\mathbf{H}$ can be written as $\mathbf{H}=\int_{\Sigma}dx^3h$, where $h=h(\mathcal{C},\mathcal{C}_a)$ takes different formulations for different deparametrization models. The diffeomorphism constraint $\mathcal{C}_a$ and the  Hamiltonian constraint $\mathcal{C}$ are given by
\begin{equation}\label{VC}
 \mathcal{C}_a:=E^b_iF^i_{ab},
\end{equation}
and
\begin{eqnarray}\label{SC}
\mathcal{C}:=\frac{E^a_i E^b_j}{\sqrt{|\det{(E)}|}}({\epsilon^{ij}}_kF^k_{ab}-2(1+\beta^2)K^i_{[a}K^j_{b]}),
\end{eqnarray}
respectively, where $F_{ab}^i=\partial_aA_b^i-\partial_bA_a^i+\epsilon_{ijk}A_a^jA_b^k$ is the curvature of $A_a^i$.
Equivalently, the Hamiltonian can also be given by \cite{Alesci:2015wla,Assanioussi:2015gka}
\begin{equation}\label{SC2}
\mathcal{C}:=-\frac{1}{\beta^2}\frac{E^a_i E^b_j}{\sqrt{|\det{(E)}|}}{\epsilon^{ij}}_kF^k_{ab}-(1+\frac{1}{\beta^2})\sqrt{|\det{(E)}|}R\approx0,
\end{equation}
where  $\sqrt{|\det{(E)}|}R:=-\sqrt{|\det{(E)}|}R_{ab}^j\epsilon_{jkl}e^{ak}e^{bl}$ is the  densitized scalar curvature of the spatial metric $q_{ab}$, with  $R_{ab}^j:=2\partial_{[a}\Gamma_{b]}^j+\epsilon^j_{\ kl}\Gamma_{a}^k\Gamma^l_b$ \cite{thiemann2008modern}.

The loop quantization of the $SU(2)$ connection formulation of GR leads to a kinematical Hilbert space $\mathcal{H}$, which can be regarded as a union of the Hilbert spaces $\mathcal{H}_\gamma=L^2((SU(2))^{|E(\gamma)|},d\mu_{\text{Haar}}^{|E(\gamma)|})$ on all possible graphs $\gamma$,  where $|E(\gamma)|$ denotes the number of independent edges of $\gamma$ and $d\mu_{\text{Haar}}$ denotes the Haar measure on $SU(2)$. In this sense, on each given $\gamma$ there is a discrete phase space $(T^\ast SU(2))^{|E(\gamma)|}$, which is coordinatized by the basic discrete variables---holonomies and fluxes. The holonomy of $A_a^i$ along an edge $e\in\gamma$ is defined by
 \begin{equation}
h_e[A]:=\mathcal{P}\exp(\int_eA)=1+\sum_{n=1}^{\infty}\int_{0}^1dt_n\int_0^{t_n}dt_{n-1}...\int_0^{t_2} dt_1A(t_1)...A(t_n),
 \end{equation}
 where $A(t)=A_a^i(t)\dot{e}^a(t)\tau_i$, and $\tau_i=-\frac{\mathbf{i}}{2}\sigma_i$ with $\sigma_i$ being the Pauli matrices.
There are two versions for the gauge covariant flux of $E^b_j$ through the 2-face dual to edge $e\in \gamma$ \cite{Thomas2001Gauge,Zhang:2021qul}. The flux $F^i_e$ (or denoted by $ F^i(e)$) in the perspective of source point of $e$ is defined by
 \begin{equation}\label{F111}
 F^i_e:=\frac{2}{\beta }\text{tr}\left(\tau^i\int_{S_e}\epsilon_{abc}h(\rho^s_e(\sigma))E^{cj}(\sigma)\tau_jh(\rho^s_e(\sigma)^{-1})\right),
 \end{equation}
 where $S_e$ is the 2-face in the dual lattice $\gamma^\ast$ of $\gamma$, $\rho^s(\sigma): [0,1]\rightarrow \Sigma$ is a path connecting the source point $s_e\in e$ to $\sigma\in S_e$ such that $\rho_e^s(\sigma): [0,\frac{1}{2}]\rightarrow e$ and $\rho_e^s(\sigma): [\frac{1}{2}, 1]\rightarrow S_e$.  Similarly, the corresponding flux $\tilde{F}^i_e$ (or denoted by $\tilde{F}^i(e)$) in the perspective of target point of $e$ is defined by
  \begin{equation}\label{F222}
 \tilde{F}^i_e:=-\frac{2}{\beta }\text{tr}\left(\tau^i\int_{S_e}\epsilon_{abc}h(\rho^t_e(\sigma))E^{cj}(\sigma)\tau_jh(\rho^t_e(\sigma)^{-1})\right),
 \end{equation}
 where $\rho^t(\sigma): [0,1]\rightarrow \Sigma$ is a path connecting the target point $t_e\in e$ to $\sigma\in S_e$ such that $\rho_e^t(\sigma): [0,\frac{1}{2}]\rightarrow e$ and $\rho_e^t(\sigma): [\frac{1}{2}, 1]\rightarrow S_e$. It is easy to see that one has the relation
 \begin{equation}\label{F333}
 \tilde{F}^i_e\tau_i=-h_e^{-1} {F}^i_e\tau_ih_e.
 \end{equation}
 The non-vanishing Poisson brackets among the holonomy and fluxes read
 \begin{eqnarray}\label{hp1220}
&&\{{{h}}_e[{A}],{{F}}^i_{e'}\}= -\delta_{e,e'}{\kappa}\tau^i{{h}}_e[{A}],\quad \{{{h}}_e[{A}],\tilde{{F}}^i_{e'}\}= \delta_{e,e'}{\kappa}{{h}}_e[{A}]\tau^i,\\\nonumber
&&\{{{F}}^i_{e},{{F}}^j_{e'}\}= -\delta_{e,e'}{\kappa}{\epsilon^{ij}}_k{{F}}^k_{e'},\quad \{\tilde{{F}}^i_{e},\tilde{{F}}^j_{e'}\}= -\delta_{e,e'}{\kappa}{\epsilon^{ij}}_k\tilde{{F}}^k_{e'}.
\end{eqnarray}

The basic operators in $\mathcal{H}_\gamma$ is given by promoting the basic discrete variables as operators. The resulting holonomy and flux operators act on cylindrical functions $f_\gamma(A)=f_\gamma(h_{e_1}[A],...,h_{e_{|E(\gamma)|}}[A])$ in $\mathcal{H}_\gamma$ as
\begin{equation}
\hat{h}_e[A]f_\gamma(A)=h_e[A]f_\gamma(A),
\end{equation}
\begin{equation}
\hat{F}^i_ef_\gamma(h_{e_1}[A],...,h_{e}[A],...,h_{e_{|E(\gamma)|}}[A]=\mathbf{i}\kappa\hbar\frac{d}{d\lambda} f_\gamma\left.\left(h_{e_1}[A],...,e^{\lambda\tau^i}h_{e}[A],...,h_{e_{|E(\gamma)|}}[A]\right)\right|_{\lambda=0},
\end{equation}
\begin{equation}
\hat{\tilde{F}}^i_ef_\gamma(h_{e_1}[A],...,h_{e}[A],...,h_{e_{|E(\gamma)|}}[A]=-\mathbf{i}\kappa\hbar\frac{d}{d\lambda} f_\gamma\left.\left(h_{e_1}[A],...,h_{e}[A]e^{\lambda\tau^i},...,h_{e_{|E(\gamma)|}}[A]\right)\right|_{\lambda=0}.
\end{equation}
Two spatial geometric operators in $H_\gamma$ are worth being mentioned here. The first one is the oriented area operator defined as $\beta \hat{F}^i_e$ (or $\beta \hat{\tilde{F}}^i_e$), whose module length $|\beta \hat{F}_e|:=\sqrt{\beta^2\hat{F}^i(e)\hat{F}_i(e)}$ represents the area of the 2-face dual to $e$ and direction represents the ingoing normal direction of $S_e$ in the perspective of the source (or target) point of $e$. As a remarkable prediction of LQG, the module length of the oriented area operator takes the following discrete spectrum \cite{Han2005FUNDAMENTAL},
 \begin{equation}\label{eigenp0}
 \text{Spec}(|\beta \hat{F}_e|)=\{\beta \kappa\hbar\sqrt{j(j+1)}|j\in\frac{\mathbb{N}}{2}\}.
 \end{equation}
 The second important spatial geometric operator is the volume operator of a compact region $D\subset \Sigma$, which is defined as \cite{Ashtekar:1997fb}
\begin{equation}\label{Vdef}
\hat{V}_D:=\sum_{v\in V(\gamma)\cap D}\hat{V}_v=\sum_{v\in V(\gamma)\cap D}\sqrt{|\hat{Q}_v|},
\end{equation}
where $V(\gamma)$ denotes the set of vertices of $\gamma$, and
\begin{equation}
\hat{Q}_v:=\frac{1}{8}(\beta )^3\sum_{\{e_I,e_J,e_K\}\subset E(\gamma)}^{e_I\cap e_J\cap e_K=v}\epsilon_{ijk}\epsilon^{IJK}\hat{F}^i(v,e_I)\hat{F}^j(v,e_J)\hat{F}^k(v,e_K),
\end{equation}
where $\epsilon^{IJK}=\text{sgn}[\det(e_I\wedge e_J\wedge e_K)]$, $\hat{F}^i(v,e)=\hat{F}^i(e)$ if $s(e)=v$ and $\hat{F}^i(v,e)=-\hat{\tilde{F}}^i(e)$ if $t(e)=v$.

The Gaussian constraint operator can be well defined in $\mathcal{H}_\gamma$ as well as in $\mathcal{H}$, which generates $SU(2)$ gauge transformations of the cylindrical functions. The solution space of the quantum Gaussian constraint is composed by the gauge invariant spin-network states, and it is also  the physical Hilbert space in the deparametriztion models of LQG. Correspondingly, the quantum dynamics of LQG is governed by the true physical Hamiltonian operator $\hat{\mathbf{H}}$ \cite{Giesel:2007wn,Thiemann:2023zjd,Brown:1994py,PhysRevD.43.419,Domagala:2010bm}. To adapt the further constructions in this article, let us consider the Hamiltonian operator in $\mathcal{H}_\gamma$ with $\Gamma$ being a cubic graph. Following the classical expression \eqref{SC} and the construction of Giesel and Thiemann \cite{Thiemann:1996at,Giesel:2007wn}, the quantum Hamiltonian consists of the so-called Euclidean part $\hat{\mathcal{C}}_E$ and the Lorentzian part $\hat{\mathcal{C}}_L$ as the quantization of $\mathcal{C}$, which reads
\begin{equation}\label{scalarcons}
\hat{\mathcal{C}}=\hat{\mathcal{C}}_E+(1+\beta^2)\hat{\mathcal{C}}_L.
\end{equation}
 For the special model of non-graph-changing and cubic graph $\gamma$, the Euclidean part is defined as
\begin{equation}
\hat{\mathcal{C}}_E=\frac{-4}{\mathbf{i}\beta \kappa\hbar}\sum_{v\in \gamma}\sum^{e_I\cap e_J\cap e_K=v}_{e_I,e_J,e_K\in \gamma}\epsilon^{IJK}\text{tr}(h_{\alpha_{IJ}}h_{e_K}[\hat{V}_v,h^{-1}_{e_K}]),
\end{equation}
where $e_I,e_J, e_K$ have been re-oriented to be outgoing at $v$, $\epsilon^{IJK}=\text{sgn}[\det(e_I\wedge e_J\wedge e_K)]$, $\alpha_{IJ}$ is the minimal loop around a plaquette containing $e_I$ and $e_J$ \cite{Han:2020chr,Giesel_2007}, which begins at $v$ via $e_I$ and gets back to $v$ through $e_J$. With the same notations, the Lorentzian part is given by
\begin{eqnarray}
&&\hat{\mathcal{C}}_L\\\nonumber
&=&\frac{8}{\mathbf{i}\beta^7(\kappa\hbar)^5}\sum_{v\in \gamma}\sum^{e_I\cap e_J\cap e_K=v}_{e_I,e_J,e_K\in\gamma}\epsilon^{IJK}\text{tr}\left([h_{e_I},[\hat{V}_v,\hat{C}_E(1)]]h^{-1}_{e_I} [h_{e_J},[\hat{V}_v,\hat{C}_E(1)]]h^{-1}_{e_J}[h_{e_K},\hat{V}_v]h_{e_K}^{-1}\right).
\end{eqnarray}
Another proposal for the Hamiltonian is given by Alesci-Assanioussi-Lewandowski-Makinen
(AALM) based on the classical expression \eqref{SC2}, which is constituted by \cite{Alesci:2015wla,Assanioussi:2015gka}
\begin{equation}\label{scalarcons2}
\hat{\mathcal{C}}=-\frac{1}{\beta^2}\hat{\mathcal{C}}_E-(1+\frac{1}{\beta^2})\hat{\tilde{R}},
\end{equation}
where the smeared spatial curvature operator $\hat{\tilde{R}}$ is the quantization of the integral
\begin{equation}
\tilde{R}=\int_{\Sigma} dx \sqrt{|\det{(E)}|}R(x).
\end{equation}

As we have mentioned in introduction,  the previous constructions of the smeared spatial curvature operator $\hat{\tilde{R}}$ encounter kinds of issues \cite{Alesci:2015wla,Assanioussi:2015gka,Lewandowski:2021iun,Lewandowski:2022xox}, so that we would like to consider the twisted-geometry construction of this operator. In next subsection, we will start to introduce the details of twisted geometry parametrization of the holonomy-flux phase space.
\subsection{Twisted geometric parametrization of $SU(2)$ holonomy-flux phase space}
As mentioned before, quantum theory on a graph $\gamma$ is completely determined by the Hilbert space $\mathcal{H}_\gamma$ constructed on $\gamma$ and the basic holonomy and flux operators defined in $\mathcal{H}_\gamma$. The inherent holonomy-flux phase space associated with $\gamma$ is coordinatized by classical holonomy and flux variables. The holonomy-flux variables capture the discrete geometry information of the dual lattice of $\gamma$, which can be explained by the so-called twisted geometry \cite{Rovelli:2010km,Freidel:2010bw,PhysRevD.103.086016,Long:2023ivt}.
The following is a brief introduction of this parametrization.

From now on we will focus on a graph $\gamma$ whose dual lattice gives a partition of $\sigma$ consisting of 3-dimensional polytopes, and the elementary edge $e\in\gamma$ refers to such kind of edge which passes through only one 2-dimensional face in the dual lattice of $\gamma$. The discrete phase space related to the give graph $\gamma$ is given by $\times_{e\in \gamma}T^\ast SU(2)_e$, where $e$ is the elementary edges of $\gamma$ and $T^\ast SU(2)_e\cong (SU(2)\times su(2))_e\ni(h_e,p_e^i)$, with $p_e^i:=\frac{F_e^i}{a^2}$ being the dimensionless
flux and $a$ being a constant with the dimension of length. The space $\times_{e\in \gamma}T^\ast SU(2)_e$ is equipped with the symplectic 1-form
\begin{equation}\label{sym1}
  \Theta_{\gamma}=\sum_{e\in\gamma}\text{Tr}(p_e^i\tau_idh_eh_e^{-1}),
\end{equation}
where $\text{Tr}(XY):=-2\text{tr}_{1/2}(XY)$ with $X,Y\in su(2)$.
Without loss of generality, we can first focus on the space $T^\ast SU(2)_e$ related to one single elementary edge $e\in\gamma$.
This space can be parametrized by using the so-called twisted geometry variables
 \begin{equation}
 (V_e,\tilde{V}_e,\xi_e, \eta_e)\in P_e:=S^2_e\times S^2_e\times T^\ast S^1_e,
 \end{equation}
 where $\eta_e\in\mathbb{R}$,  $\xi_e\in [-2\pi,2\pi)$, and
  \begin{equation}
  V_e:=V_e^i\tau_i, \   \ \  \tilde{V}_e:=\tilde{V}^i_e\tau_i,
  \end{equation}
with $S^2_e$ being the space of unit vectors $V_e^i$ or $\tilde{V}^i_e$. To capture the intrinsic curvature, we specify one pair of the $SU(2)$ valued Hopf sections $n_e:=n_e(V_e)$ and $\tilde{n}_e:=\tilde{n}_e( \tilde{V}_e)$ which satisfies $V^i_e\tau_i=n_e\tau_3n_e^{-1}$ and $\tilde{V}^i_e\tau_i=-\tilde{n}_e\tau_3\tilde{n}_e^{-1}$. Then, the parametrization associated with each edge is given by the map
\begin{eqnarray}\label{para}
(V_e,\tilde{V}_e,\xi_e,\eta_e)\mapsto(h_e,p^i_e)\in T^\ast SU(2)_e:&& p^i_e\tau_i=\eta_e V_e=\eta_en_e(V_e)\tau_3n_e(V_e)^{-1}\\\nonumber
&&h_e=n_e(V_e)e^{\xi_e\tau_3}\tilde{n}_e(\tilde{V}_e)^{-1}.
\end{eqnarray}
One should note that this map is a two-to-one double cover.
In other words, under the map \eqref{para}, the two points $(V_e,\tilde{V}_e,\xi_e,\eta_e)$ and $(-V_e,-\tilde{V}_e,-\xi_e,-\eta_e)$ are mapped to the same point $(h_e,p^i_e)\in T^\ast SU(2)_e$.
Hence, by selecting either branch among the two signs related by a $\mathbb{Z}_2$ symmetry, one can establish a bijection map in the region $\eta_e\neq0$.
 Now we can get back to the discrete phase space of  LQG on the whole graph $\gamma$, which is just the Cartesian product of the discrete phase space on every single edge of $\gamma$. The twisted geometry parametrization of the discrete phase space on one copy of the edge can be directly generalized to that of the whole graph $\gamma$, with the twisted geometry parameters $(V_e,\tilde{V}_e,\xi_e, \eta_e)$ taking the interpretation of the discrete geometry describing the dual lattice of $\gamma$. Let us explain this explicitly as follows. We first note that $\eta _e V^i_e$ and $\eta _e \tilde{V}^i_e$ represent the area-weighted outward normal vectors of the 2-dimensional face dual to $e$ in the perspective of the source and target points of $e$ respectively, with $\eta _e$ being the dimensionless area of the 2-dimensional face dual to $e$. Then, the holonomy $h_e=n_e(V_e)e^{\xi_e\tau_3}\tilde{n}^{-1}_e(\tilde{V}_e)$ rotates the inward normal $-\eta _e\tilde{V}^i_e$ of the 2-dimensional dual to $e$ in the perspective of the the target point of $e$, into the outward normal $\eta _e{V}^i_e$ of the 2-dimensional face  dual to $e$ in the perspective of the source point of $e$ by
  \begin{equation}
  \tilde{V}_e=-h_e^{-1}V_eh_e,
  \end{equation}
  wherein $n_e(V_e)$ and $\tilde{n}_e(\tilde{V}_e)$ capture the contribution of intrinsic curvature, and $e^{\xi_e\tau_3}$ captures the contribution of extrinsic curvature to this rotation. Now, we have the twisted geometry parameter space $P_\gamma=\times_{e\in\gamma}P_e, P_e:=S^2_e\times S^2_e\times T^\ast S^1_e$ associated to $\gamma$, equipped with the symplectic 1-form
  \begin{equation}\label{sym2}
  \Theta_{P_\gamma}=\sum_{e\in\gamma}\eta_e\text{Tr}(V_edn_en_e^{-1})+\eta_ed\xi_e +\eta_e\text{Tr}(\tilde{V}_ed\tilde{n}_e\tilde{n}_e^{-1}).
\end{equation}
It has been checked that the map \eqref{para} provides an invertible symplectomorphism between the phase
space $\times_{e\in \gamma}T^\ast SU(2)_e$ with symplectic 1-form \eqref{sym1} and $P_\gamma$ with symplectic 1-form \eqref{sym2} in the region $\eta_e\neq0$ (up to the $\mathbb{Z}_2$ symmetry).
Then, beginning with the twisted geometry parameter space $P_\gamma$, one can proceed the gauge reduction with respect to the discrete Gauss constraint
 \begin{equation}
G_v:= -\sum_{e,s(e)=v}{p}^i_e\tau_i+\sum_{e,t(e)=v}p^i_eh_e^{-1}\tau_ih_e=0.
 \end{equation}
{ The finite gauge transformation $\{g_v|v\in\gamma\}$ generated by Gauss constraint can be given as
 \begin{equation}
 h_e\mapsto g_{s(e)}h_eg_{t(e)}^{-1},\quad {p}^i_e\tau_i\mapsto {p}^i_e g_{s(e)}\tau_i g_{s(e)}^{-1}.
 \end{equation}
 Correspondingly, the gauge transformation of the twisted geometry variables is given by
 \begin{eqnarray}\label{gaugetransV}
  {V}^i_e\tau_i\mapsto V_e(g_{s(e)}):= {V}^i_e g_{s(e)}\tau_i g_{s(e)}^{-1},  &&  \tilde{V}^i_e\tau_i\mapsto \tilde{V}_e(g_{t(e)}):=\tilde{V}^i_e g_{t(e)}\tau_i g_{t(e)}^{-1},\\\nonumber
  \xi_e\mapsto   \xi_e+\xi_{g_{s(e)}}-\xi_{g_{t(e)}}, &&  \eta_e\mapsto\eta_e,
 \end{eqnarray}
 where $\xi_{g_{s(e)}}$ and $\xi_{g_{t(e)}}$ are determined by
 \begin{eqnarray}\label{xige}
 && g_{s(e)}n_e=n_e(V_e(g_{s(e)}))e^{\xi_{g_{s(e)}}\tau_3},\\\nonumber
  &&g_{t(e)}\tilde{n}_e=\tilde{n}_e(\tilde{V}_e(g_{t(e)}))e^{\xi_{g_{t(e)}}\tau_3}
 \end{eqnarray}
  respectively.}
The gauge reduction with respect to the discrete Gauss constraint leads to the reduced phase space
 \begin{equation}
{H}_\gamma:={P}_\gamma/\!/SU(2)^{|V(\gamma)|}=\left(\times_{e\in\gamma} T^\ast S_e^1\right)\times \left(\times_{v\in\gamma} \mathfrak{P}_{\vec{\eta}_v}\right)
\end{equation}
 with $|V(\gamma)|$ being the number of the vertices in $\gamma$. The space $\mathfrak{P}_{\vec{\eta}_v}$ is the shape space of the 3-dimensional polyhedra dual to $v$ \cite{PhysRevD.83.044035,Long:2020agv}, which is given by
 \begin{equation}
 \mathfrak{P}_{\vec{\eta}_v}:=\{(V_{e_1},...,V_{e_{n_v}})\in \times_{e\in\{e_v\}}S^{2}_{e}| G_{v}=0 \}/SU(2),
 \end{equation}
where we re-oriented the edges linked to $v$ to be out-going at $v$ without loss of generality and $\{e_v\}$ represents the set of edges beginning at $v$ with $n_v$ being the number of elements in $\{e_v\}$. It has been shown that the twisted geometry described by the parameters in ${H}_\gamma$ is consistent with the Regge geometry on the spatial 3-manifold $\sigma$ if the shape-matching condition of 2-dimensional faces in the gluing process of the 3-dimensional polyhedra is satisfied \cite{PhysRevD.83.044035}.

\section{Holonomy of spin connection and spatial scalar curvature in twisted geometry}\label{sec3}
\subsection{Holonomy of spin connection in terms of twisted geometry}
\begin{figure}[h]
 \includegraphics[scale=0.12]{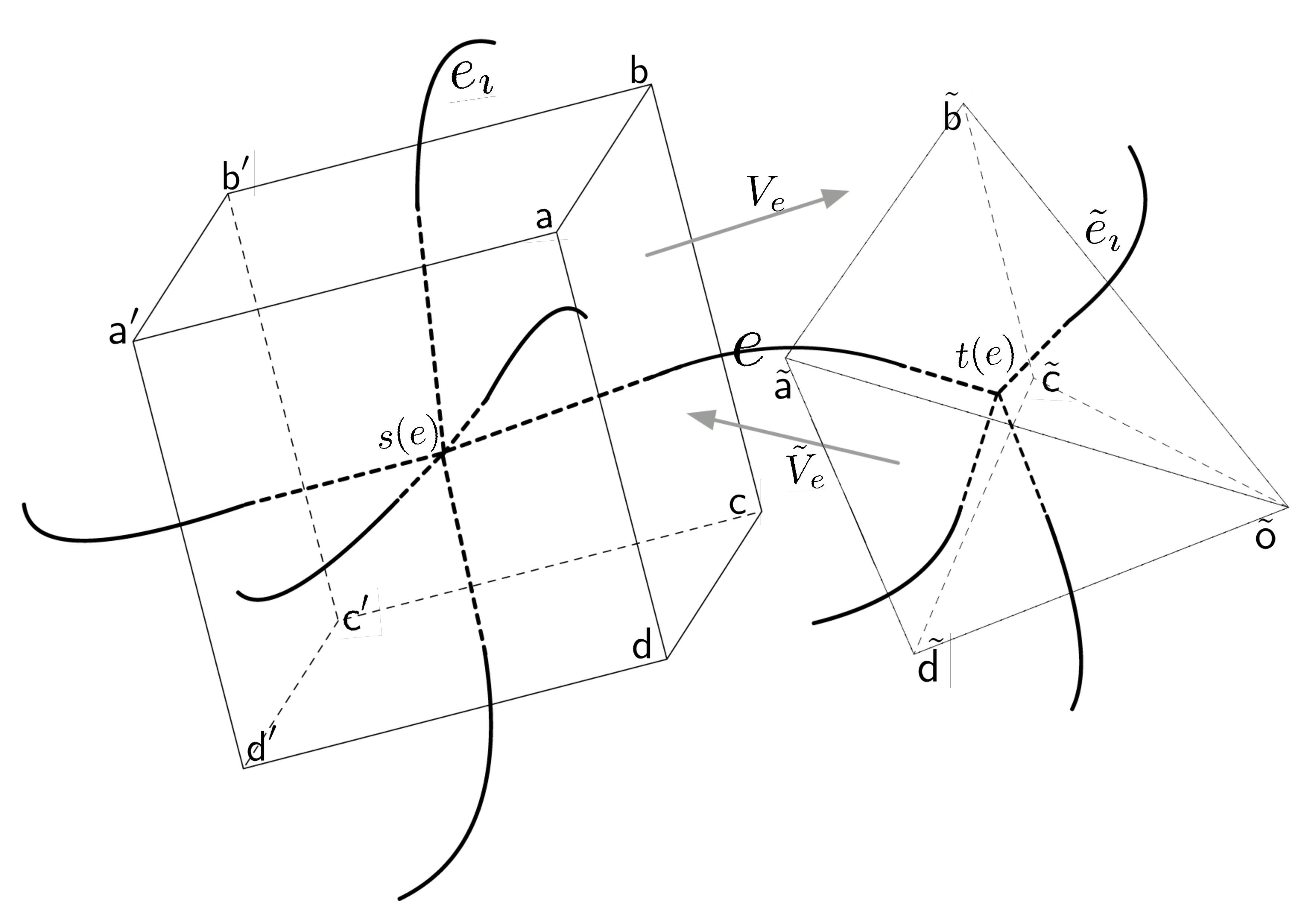}
\caption{This figure shows the geometric interpretation of each factor in the expression $h_e^{\Gamma}=n_e(V_e)e^{\zeta_e \tau_3}\tilde{n}_e(\tilde{V}_e)^{-1}$. The source point $s(e)$ of edge $e$ is dual to a hexahedron $\textsf{a}\textsf{b}\textsf{c}\textsf{d}\textsf{a}'\textsf{b}'\textsf{c}'\textsf{d}'$ and the target point $t(e)$ of edge $e$ is dual to a quadrangular pyramid $\tilde{\textsf{a}}\tilde{\textsf{b}}\tilde{\textsf{c}}\tilde{\textsf{d}}\tilde{\textsf{o}}$, and edge $e$ is dual to the two faces $\textsf{a}\textsf{b}\textsf{c}\textsf{d}$ and $\tilde{\textsf{a}}\tilde{\textsf{b}}\tilde{\textsf{c}}\tilde{\textsf{d}}$, with $\textsf{abcd}$ and $\tilde{\textsf{a}}\tilde{\textsf{b}}\tilde{\textsf{c}}\tilde{\textsf{d}}$ belonging to the hexahedron $\textsf{a}\textsf{b}\textsf{c}\textsf{d}\textsf{a}'\textsf{b}'\textsf{c}'\textsf{d}'$ and quadrangular pyramid $\tilde{\textsf{a}}\tilde{\textsf{b}}\tilde{\textsf{c}}\tilde{\textsf{d}}\tilde{\textsf{o}}$ respectively. The spin connection holonomy  $h_e^{\Gamma}=n_e(V_e)e^{\zeta_e \tau_3}\tilde{n}_e(\tilde{V}_e)^{-1}$ tells us how to glue the hexahedron $\textsf{a}\textsf{b}\textsf{c}\textsf{d}\textsf{a}'\textsf{b}'\textsf{c}'\textsf{d}'$ and quadrangular pyramid $\tilde{\textsf{a}}\tilde{\textsf{b}}\tilde{\textsf{c}}\tilde{\textsf{d}}\tilde{\textsf{o}}$ by matching the two faces $\textsf{a}\textsf{b}\textsf{c}\textsf{d}$ and $\tilde{\textsf{a}}\tilde{\textsf{b}}\tilde{\textsf{c}}\tilde{\textsf{d}}$. Specifically,  $n_e(V_e)^{-1}$ rotates the hexahedron $\textsf{a}\textsf{b}\textsf{c}\textsf{d}\textsf{a}'\textsf{b}'\textsf{c}'\textsf{d}'$ to ensure that the outward and unit normal vector $V^i_e$ of the face $\textsf{a}\textsf{b}\textsf{c}\textsf{d}$ equals to the unit vector $\delta_3^i$, while $\tilde{n}_e(\tilde{V}_e)^{-1}$ rotates the quadrangular pyramid $\tilde{\textsf{a}}\tilde{\textsf{b}}\tilde{\textsf{c}}\tilde{\textsf{d}}\tilde{\textsf{o}}$ to ensure that the outward and unit normal vector $\tilde{V}^i_e$ of the face $\tilde{\textsf{a}}\tilde{\textsf{b}}\tilde{\textsf{c}}\tilde{\textsf{d}}$ to the unit vector $-\delta_3^i$. Moreover, with $V_e^i=-\tilde{V}_e^i=\delta_3^i$, $e^{\zeta_e \tau_3}$ rotates the quadrangular pyramid $\tilde{\textsf{a}}\tilde{\textsf{b}}\tilde{\textsf{c}}\tilde{\textsf{d}}\tilde{\textsf{o}}$ to ensure that an edge of the face $\tilde{\textsf{a}}\tilde{\textsf{b}}\tilde{\textsf{c}}\tilde{\textsf{d}}$ is parallel to its corresponding edge in the face $\textsf{a}\textsf{b}\textsf{c}\textsf{d}$. The two edges from the faces $\textsf{a}\textsf{b}\textsf{c}\textsf{d}$ and $\tilde{\textsf{a}}\tilde{\textsf{b}}\tilde{\textsf{c}}\tilde{\textsf{d}}$, which will be parallel to each other with the rotation of $\tilde{\textsf{a}}\tilde{\textsf{b}}\tilde{\textsf{c}}\tilde{\textsf{d}}\tilde{\textsf{o}}$ by $e^{\zeta_e \tau_3}$, can be determined by choosing a minimal loop containing $e$. For instance, one can chooses the minimal loop $\square_\imath$ containing $e$, $e_\imath$ and $\tilde{e}_\imath$, and then edges $\tilde{\textsf{a}}\tilde{\textsf{b}}$ and $\textsf{a}\textsf{b}$ dual to $\square_\imath$ will be parallel to each other with the rotation of $\tilde{\textsf{a}}\tilde{\textsf{b}}\tilde{\textsf{c}}\tilde{\textsf{d}}\tilde{\textsf{o}}$ by $e^{\zeta_e \tau_3}$.  Thus, $e^{\zeta_e \tau_3}$ is dependent on the choice of the minimal loop containing $e$.  If the shape of the faces $\textsf{a}\textsf{b}\textsf{c}\textsf{d}$ and $\tilde{\textsf{a}}\tilde{\textsf{b}}\tilde{\textsf{c}}\tilde{\textsf{d}}$ are matched to each other, the twisted geometry on $e$ will be referred as to the shape-matching twisted geometry on $e$. Then, it is easy to see that $e^{\zeta_e \tau_3}$ is independent of the choice the minimal loop for the shape-matching twisted geometry on $e$.}
\label{fig:label1}
\end{figure}

{The holonomy $h_e^{\Gamma}$ of the spin connection  can be factorized out from the holonomy of Ashtekar connection based on the twisted geometry parametrization of holonomy-flux phase space as follows  \cite{Rovelli:2010km}. First, one can choose a gauge at $s(e)$ and $t(e)$ to ensure $V_e=-\tilde{V}_e$, and then the holonomy $h_e$ takes the formulation $h_e=n_e(V_e)e^{\xi_e\tau_3}n_e(V_e)^{-1}=e^{\xi_eV_e}$ in this gauge. Also, it is know that the contribution of extrinsic curvature to the holonomy $h_e$ takes the formulation $e^{\beta\tilde{\theta}_eV_e}$ in this gauge \cite{Rovelli:2010km}. Then, the holonomy $h_e^{\Gamma}$ of the spin connection in this gauge can be given by $h_e^{\Gamma}=e^{\zeta_e V_e}\equiv e^{(\xi_e-\beta\tilde{\theta}_e) V_e}$. Further, by recalling the gauge transformation \eqref{gaugetransV} and carrying out a gauge transformation at $t(e)$ to release the gauge $V_e=-\tilde{V}_e$ ,} one can immediately get the general formulation for the holonomy $h_e^{\Gamma}$ of the spin connection in twisted geometry, which reads
\begin{equation}\label{hnn}
h_e^{\Gamma}=n_e(V_e)e^{\zeta_e \tau_3}\tilde{n}_e(\tilde{V}_e)^{-1}.
\end{equation}
{Moreover,  by following the gauge transformation of $\{(\xi_e,V_e, \tilde{V}_e)|e\in\gamma\}$ given by Eqs.\eqref{gaugetransV}, it is easy to see that the holonomy $h^{\Gamma}$ of spin connection is transformed as
 \begin{equation}\label{gaugetranshgamma}
 h^{\Gamma}_e\mapsto g_{s(e)}h^{\Gamma}_eg_{t(e)}^{-1}
 \end{equation}
for the finite gauge transformation $\{g_v|v\in\gamma\}$ generated by Gauss constraint. Correspondingly,
 $\zeta_e$ is transformed as
\begin{equation}\label{zetatrans}
\zeta_e\mapsto   \zeta_e+\xi_{g_{s(e)}}-\xi_{g_{t(e)}}
\end{equation}
with $\xi_{g_{s(e)}}$ and $\xi_{g_{t(e)}}$ being given by Eq.\eqref{xige}.
Now,  it is clear to explain the geometric interpretation of $(n_e(V_e),\tilde{n}_e(\tilde{V}_e),\zeta_e)$ in $h_e^{\Gamma}$. As shown in Fig.\ref{fig:label1}, $n_e(V_e)$ and $\tilde{n}_e(\tilde{V}_e)$ give the rotation of the source and target polyhedrons of $e$ to ensure $V_e=-\tilde{V}_e=\tau_3$, and $e^{\zeta_e \tau_3}$ gives the rotation of the source or target polyhedron around the normal vectors $V_e=-\tilde{V}_e=\tau_3$ to ensure these two polyhedrons being aligned at the 2-face dual to $e$, see the details in Fig.\ref{fig:label1}. 

The geometric interpretation of $h_e^{\Gamma}$ helps us to express
the holonomy of the spin connection   in terms of fluxes, let us show it explicitly as follows.}
Notice that
$n_e(V_e)$ is a function of $V_e^i$, which reads
\begin{equation}
n_e(V_e)=\cos(\theta_e/2) \mathbb{I} -\mathbf{i}\sin(\theta_e/2)\sigma(V_e),
\end{equation}
where $\cos\theta_e:=V_e^3=V_e^i \delta_3^j\delta_{ij}$, $\sigma(V_e):=\frac{1}{\sin\theta_e}\delta_3^iV_e^j\epsilon_{ijk}\sigma^k$ with $\sin\theta_e=|\delta_3^iV_e^j\epsilon_{ijk}|$ and $0\leq\theta_e<\pi$. One can further express $n_e(V_e)$ as
\begin{equation}
n_e(V_e)=\sqrt{\frac{1+V_e^3}{2}} \mathbb{I} -\mathbf{i}\sqrt{\frac{1}{2(1+V_e^3)}}(V_e^1\sigma^2-V_e^2\sigma^1).
\end{equation}
and $n_e(V_e)^{-1}$ as
 \begin{equation}
n_e(V_e)^{-1}=\cos(\theta_e/2) \mathbb{I} +\mathbf{i}\sin(\theta_e/2)\sigma(V_e)=\sqrt{\frac{1+V_e^3}{2}} \mathbb{I} +\mathbf{i}\sqrt{\frac{1}{2(1+V_e^3)}}(V_e^1\sigma^2-V_e^2\sigma^1).
\end{equation}
Similarly, $\tilde{n}_e(\tilde{V}_e)$ is a function of $\tilde{V}_e^i$ given by
\begin{equation}
\tilde{n}_e(\tilde{V}_e)=\cos(\tilde{\theta}_e/2) \mathbb{I} +\mathbf{i}\sin(\tilde{\theta}_e/2)\sigma(\tilde{V}_e),
\end{equation}
where $\cos\tilde{\theta}_e:=-\tilde{V}_e^3=-\tilde{V}_e^i \delta_3^j\delta_{ij}$, $\sigma(\tilde{V}_e):=\frac{1}{\sin\tilde{\theta}_e}\delta_3^i\tilde{V}_e^j\epsilon_{ijk}\sigma^k$ with $ \sin\tilde{\theta}_e=|\delta_3^i\tilde{V}_e^j\epsilon_{ijk}|$ and $0\leq\tilde{\theta}_e<\pi$.
We also express $\tilde{n}_e(\tilde{V}_e)$ and $\tilde{n}_e(\tilde{V}_e)^{-1}$ as
\begin{equation}
\tilde{n}_e(\tilde{V}_e)=\sqrt{\frac{1-\tilde{V}_e^3}{2}}\mathbb{I} +\mathbf{i}\sqrt{\frac{1}{2(1-\tilde{V}_e^3)}}(\tilde{V}_e^1\sigma^2-\tilde{V}_e^2\sigma^1),
\end{equation}
and
\begin{equation}
\tilde{n}_e(\tilde{V}_e)^{-1}=\sqrt{\frac{1-\tilde{V}_e^3}{2}}\mathbb{I} -\mathbf{i}\sqrt{\frac{1}{2(1-\tilde{V}_e^3)}}(\tilde{V}_e^1\sigma^2-\tilde{V}_e^2\sigma^1),
\end{equation}
respectively.

Now, let us turn to express $e^{\zeta_e \tau_3}$ in terms of fluxes. { By rotating the source and target polyhedrons of $e$ to ensure $V_e=-\tilde{V}_e=\tau_3$,    $e^{\zeta_e \tau_3}$ gives the rotation of the source or target polyhedron around the normal vectors $V_e=-\tilde{V}_e=\tau_3$ to ensure these two polyhedrons being aligned  at the faces dual to $e$. Nevertheless, this alignment is still an undetermined notation for twisted geometry. In fact, for the Regge geometry sector of twisted geometry satisfying the shape-matching condition,  the faces dual to $e$ that are respectively from the source and target polyhedrons have identical shape, and thus the alignment is determined naturally. However, for the general twisted geometry,  the shapes of the faces dual to $e$ in  the frames of source and target polyhedrons respectively  are not necessarily identical. Thus, the strict alignment of source and target polyhedrons can not be achieved at the faces dual to $e$. Indeed, this issue has been studied in the case that both of source and target polyhedrons are tetrahedrons in Ref. \cite{PhysRevD.87.024038}, in which a relaxed alignment is proposed to construct the spin connection of the frames associated to the glued tetrahedrons  in twisted geometry, with this relaxed alignment relying on the choice of the frame on the glued triangles dual to $e$. In this article, we would like to consider another type of relaxed alignment adapted to arbitrary polyhedrons, which only  relies on one edge of each face dual to $e$ (e.g. the edge $\tilde{\textsf{a}}\tilde{\textsf{b}}$ or $\textsf{a}\textsf{b}$ in Fig.\ref{fig:label1}).  Notice that the edge used to determine the relaxed alignment is dual to  a minimal loop in $\gamma$ containing $e$,
}  thus we can choose the minimal loop $\square_\imath$ containing $e$, $e_\imath$ and $\tilde{e}_\imath$ to illustrate our construction without loss of generality, see more details in Fig.\ref{fig:label1}. Then, one can define
\begin{equation}
\check{V}_{e_\imath}^i:=-2\text{tr}(\tau^i n_e^{-1}\tau_jn_e)\breve{V}_{e_\imath}^j=\frac{\bar{V}_{e_\imath}^i-{V}_{e_\imath}^kV_{e,k}\delta_3^i} {\sqrt{1-({V}_{e_\imath}^kV_{e,k})^2}}
\end{equation}
at the source point of $e$,
where $\breve{V}_{e_\imath}^j:=\frac{{V}_{e_\imath}^j-{V}_{e_\imath}^kV_{e,k}V_{e}^j} {\sqrt{1-({V}_{e_\imath}^kV_{e,k})^2}}$, $e_\imath$ is reoriented to be started at $s(e)$, and
\begin{eqnarray}
\bar{V}_{e_\imath}^i&:=&-2\text{tr}(\tau^i n_e^{-1}\tau_jn_e){V}_{e_\imath}^j \\\nonumber &=&\frac{\delta_3^jV_e^k{\epsilon_{jk}}^i}{\sin^2\theta_e}\delta_3^{i'}V_e^{j'} \epsilon_{i'j'k'}{V}_{{e}_{\imath}}^{k'} +2\delta_3^{[i}V_e^{j]}{V}_{{e}_{\imath},j}-4\delta_3^{[i}V_e^{k]}\delta_{kk'} \delta_3^{[k'}V_e^{j]}{V}_{{e}_{\imath},j}\frac{\cos\theta_e}{\sin^2\theta_e}\\\nonumber
&=&\frac{\delta_3^jV_e^k{\epsilon_{jk}}^i}{1-(V_e^3)^2}(V_e^{1}{V}_{{e}_{\imath}}^{2}-V_e^{2} {V}_{{e}_{\imath}}^{1})+\frac{1-2(V_e^3)^2}{1-(V_e^3)^2}\delta_3^{i} V_e^{j}{V}_{{e}_{\imath},j}\\\nonumber
 &&+\frac{V_e^3}{1-(V_e^3)^2}\delta_3^{i}V_{e,k'} V_e^{k'}{V}^3_{{e}_{\imath}}+\frac{V_e^3}{1-(V_e^3)^2}V_e^{i} V_e^{j}{V}_{{e}_{\imath},j}-\frac{1}{1-(V_e^3)^2}V_e^{i} {V}^3_{{e}_{\imath}}.
\end{eqnarray}
Also, let us define
\begin{equation}
 \check{V}_{\tilde{e}_{\imath}}^i:=-2\text{tr}(\tau^i\tilde{n}_e^{-1}\tau_j\tilde{n}_e)\breve{V}_{\tilde{e}_{\imath}}^j =\frac{\bar{V}_{\tilde{e}_\imath}^i+{V}_{\tilde{e}_\imath}^k\tilde{V}_{e,k}\delta_3^i} {\sqrt{1-({V}_{\tilde{e}_\imath}^k\tilde{V}_{e,k})^2}}
\end{equation}
at the target point of $e$,
where $\breve{V}_{\tilde{e}_{\imath}}^j:=\frac{{V}_{\tilde{e}_\imath}^j-{V}_{\tilde{e}_\imath}^k\tilde{V}_{e,k}\tilde{V}_{e}^j} {\sqrt{1-({V}_{\tilde{e}_\imath}^k\tilde{V}_{e,k})^2}}$, $\tilde{e}_\imath$ is reoriented to be started at $t(e)$,  and
\begin{eqnarray}
\bar{V}_{\tilde{e}_\imath}^i&:=&-2\text{tr}(\tau^i\tilde{n}_e^{-1}\tau_j\tilde{n}_e){V}_{\tilde{e}_{\imath}}^j  \\\nonumber &=&\frac{\delta_3^j\tilde{V}_e^k{\epsilon_{jk}}^i}{\sin^2\tilde{\theta}_e}\delta_3^{i'}\tilde{V}_e^{j'} \epsilon_{i'j'k'}{V}_{\tilde{e}_{\imath}}^{k'} -2\delta_3^{[i}\tilde{V}_e^{j]}{V}_{\tilde{e}_{\imath},j}-4\delta_3^{[i}\tilde{V}_e^{k]}\delta_{kk'} \delta_3^{[k'}\tilde{V}_e^{j]}{V}_{\tilde{e}_{\imath},j}\frac{\cos\tilde{\theta}_e}{\sin^2\tilde{\theta}_e}\\\nonumber
&=&\frac{\delta_3^j\tilde{V}_e^k{\epsilon_{jk}}^i}{1-(\tilde{V}_e^3)^2}(\tilde{V}_e^{1} {V}_{\tilde{e}_{\imath}}^{2} -\tilde{V}_e^{2} {V}_{\tilde{e}_{\imath}}^{1} )+\frac{2(\tilde{V}_e^{3})^2-1}{1-(\tilde{V}_e^3)^2} \delta_3^{i} \tilde{V}_e^{j}{V}_{\tilde{e}_{\imath},j}\\\nonumber
&&-\frac{\tilde{V}_e^{3}}{1-(\tilde{V}_e^3)^2} \delta_3^{i}\tilde{V}_e^{k} \tilde{V}_{e,k}{V}_{\tilde{e}_{\imath},3} +\frac{\tilde{V}_e^{i} {V}_{\tilde{e}_{\imath}}^3}{1-(\tilde{V}_e^3)^2}-\frac{\tilde{V}_e^{i}\tilde{V}_e^{3}}{1-(\tilde{V}_e^3)^2} \tilde{V}_e^{j}{V}_{\tilde{e}_{\imath},j}.
\end{eqnarray}
Then, $e^{\zeta_e \tau_3}$ takes the formulation
\begin{equation}
e^{\zeta_e \tau_3}=\cos(\zeta_e/2)\mathbb{I}-\mathbf{i}\sin(\zeta_e/2)\sigma_3
\end{equation}
with
\begin{equation}
\cos(\zeta_e/2)=\sqrt{\frac{1+\cos\zeta_e}{2}},\quad \sin(\zeta_e/2)=\text{sgn}(\zeta_e)\sqrt{\frac{1-\cos\zeta_e}{2}},
\end{equation}
where $\zeta_e\in(-\pi,\pi]$, $\cos\zeta_e$ and $\text{sgn}(\zeta_e)$ are given by
\begin{equation}
\cos\zeta_e=\check{V}_{\tilde{e}_{\imath}}^i\check{V}_{e_\imath}^j\delta_{ij} =\frac{\bar{V}_{e_\imath}^i\bar{V}_{\tilde{e}_\imath}^j\delta_{ij}-{V}_{e_\imath}^kV_{e,k}\bar{V}_{\tilde{e}_\imath}^3 + {V}_{\tilde{e}_\imath}^k\tilde{V}_{e,k}\bar{V}_{e_\imath}^3-{V}_{\tilde{e}_\imath}^k\tilde{V}_{e,k}{V}_{e_\imath}^lV_{e,l}} {\sqrt{1-({V}_{e_\imath}^kV_{e,k})^2}\sqrt{1-({V}_{\tilde{e}_\imath}^k\tilde{V}_{e,k})^2}}
\end{equation}
and
\begin{equation}
\text{sgn}(\zeta_e)=\text{sgn}(\epsilon_{ijk}\delta_3^i\check{V}_{\tilde{e}_{\imath}}^j\check{V}_{e_\imath}^k) =\text{sgn}(\epsilon_{ijk}\delta_3^i\bar{V}_{\tilde{e}_{\imath}}^j\bar{V}_{e_\imath}^k)
\end{equation}
respectively.

Now, $h_e^\Gamma$ can be expressed as
\begin{eqnarray}\label{hgammafinal}
&& h_e^{\Gamma}=n_e(V_e)e^{\zeta_e \tau_3}\tilde{n}_e(\tilde{V}_e)^{-1} \\\nonumber
  &=& \left(\sqrt{\frac{1+V_e^3}{2}} \mathbb{I} -\mathbf{i}\sqrt{\frac{1}{2(1+V_e^3)}}(V_e^1\sigma^2-V_e^2\sigma^1)\right) \left(\cos(\zeta_e/2)\mathbb{I}-\mathbf{i}\sin(\zeta_e/2)\sigma_3\right) \\\nonumber
  &&\cdot\left(\sqrt{\frac{1-\tilde{V}_e^3}{2}}\mathbb{I} -\mathbf{i}\sqrt{\frac{1}{2(1-\tilde{V}_e^3)}}(\tilde{V}_e^1\sigma^2-\tilde{V}_e^2\sigma^1)\right)\\\nonumber
&=&w_0(\Gamma_e)\mathbb{I}+w_1(\Gamma_e)\mathbf{i}\sigma_1+w_2(\Gamma_e)\mathbf{i}\sigma_2 +w_3(\Gamma_e)\mathbf{i}\sigma_3,
\end{eqnarray}
where we defined
\begin{eqnarray}
w_0(\Gamma_e)
&:=&- \sin(\zeta_e/2)\sqrt{\frac{1}{2(1+V_e^3)}} \sqrt{\frac{1}{2(1-\tilde{V}_e^3)}} (V_e^2\tilde{V}_e^1 -V_e^1\tilde{V}_e^2)\\\nonumber
&&+\cos(\zeta_e/2)\Big(\sqrt{\frac{1-\tilde{V}_e^3}{2}}\sqrt{\frac{1+V_e^3}{2}}-\sqrt{\frac{1}{2(1+V_e^3)}} \sqrt{\frac{1}{2(1-\tilde{V}_e^3)}} (V_e^1\tilde{V}_e^1+V_e^2\tilde{V}_e^2)\Big),
\end{eqnarray}
\begin{eqnarray}
w_1(\Gamma_e)
&:=&\cos(\zeta_e/2)\Big(\sqrt{\frac{1-\tilde{V}_e^3}{2}}\sqrt{\frac{1}{2(1+V_e^3)}}V_e^2+\sqrt{\frac{1+V_e^3}{2}} \sqrt{\frac{1}{2(1-\tilde{V}_e^3)}}\tilde{V}_e^2\Big) \\\nonumber
&& +\sin(\zeta_e/2)\Big(\sqrt{\frac{1+V_e^3}{2}} \sqrt{\frac{1}{2(1-\tilde{V}_e^3)}}  \tilde{V}_e^1-\sqrt{\frac{1-\tilde{V}_e^3}{2}}\sqrt{\frac{1}{2(1+V_e^3)}} V_e^1\Big),
\end{eqnarray}
\begin{eqnarray}
w_2(\Gamma_e)&:=&-\cos(\zeta_e/2)\Big(\sqrt{\frac{1+V_e^3}{2}} \sqrt{\frac{1}{2(1-\tilde{V}_e^3)}}\tilde{V}_e^1 +\sqrt{\frac{1-\tilde{V}_e^3}{2}}\sqrt{\frac{1}{2(1+V_e^3)}}V_e^1 \Big)\\\nonumber
&&
  +\sin(\zeta_e/2)\Big(\sqrt{\frac{1+V_e^3}{2}} \sqrt{\frac{1}{2(1-\tilde{V}_e^3)}}  \tilde{V}_e^2-\sqrt{\frac{1-\tilde{V}_e^3}{2}}\sqrt{\frac{1}{2(1+V_e^3)}} V_e^2\Big)
\end{eqnarray}
and
\begin{eqnarray}
w_3(\Gamma_e) &:=& -\cos(\zeta_e/2)\sqrt{\frac{1}{2(1+V_e^3)}} \sqrt{\frac{1}{2(1-\tilde{V}_e^3)}} (-V_e^2\tilde{V}_e^1 +V_e^1\tilde{V}_e^2) \\\nonumber
&& + \sin(\zeta_e/2)\Big(-\sqrt{\frac{1-\tilde{V}_e^3}{2}}\sqrt{\frac{1+V_e^3}{2}}+\sqrt{\frac{1}{2(1+V_e^3)}} \sqrt{\frac{1}{2(1-\tilde{V}_e^3)}} (-V_e^1\tilde{V}_e^1 -V_e^2\tilde{V}_e^2) \Big).
\end{eqnarray}

It should be noted that the holonomy $h_e^{\Gamma}$ of the spin connection given by Eq.\eqref{hgammafinal} involves a minimal loop $\square_\imath$ containing $e$, $e_\imath$ and $\tilde{e}_\imath$. Indeed, as shown in Fig.\ref{fig:label1}, if the shapes of the two faces dual to $e$ are matched, we claim that the twisted geometry on $e$ satisfies the shape-matching condition, and then the holonomy $h_e^{\Gamma}$ is independent of the choice of the minimal loop $\square_\imath$.  In general, the twisted geometry cannot ensure that the shape-matching condition holds for every edge, so the holonomy $h_e^{\Gamma}$ of the spin connection is expressed with its dependence on a minimal loop containing $e$. In the next subsection we will show that there is a natural choice of the minimal loop for $h_e^{\Gamma}$ when it appears in the expression of the discrete spatial scalar curvature.

\subsection{Spatial scalar curvature in terms of twisted geometry}
The holonomy of the spin connection in terms of flux variables helps us to regularize the densitized scalar curvature $\sqrt{q}R=-\sqrt{q}R_{ab}^j\epsilon_{jkl}e^{ak}e^{bl}$ of the spatial metric $q_{ab}$, where  $R_{ab}^j:=2\partial_{[a}\Gamma_{b]}^j+\epsilon^j_{\ kl}\Gamma_{a}^k\Gamma^l_b$. Consider a cubic graph $\gamma_{\square}$ with the coordinate length of each elementary edge being given by $\mu$, the integral $\tilde{R}=\int_{\Sigma} dx \sqrt{q}R(x)$ can be given by
\begin{equation}\label{sumRv}
\tilde{R}
=\lim_{\mu\to0}\sum_{v\in \gamma_{\square}}R_{\square_v}
\end{equation}
with $R_{\square_v}$ being given by
\begin{equation}\label{Rv1}
R_{\square_v}=\frac{(\beta a^2)^2}{2}\sum_{\square_{e_I,e_J}}\frac{p^i_{e_I}p^j_{e_J}}{V(v)}\epsilon_{ijk}\text{tr} (\tau^kh^{\Gamma}_{\square_{e_I,e_J}})
\end{equation}
where $\square_{e_I,e_J}$ is a minimal loop containing edges $e_I$ and $e_J$, which begins at $v$ via $e_I$ and gets back to $v$ through $e_J$. The second ``$=$'' in Eq.\eqref{sumRv} is from the continuum limit  $\sqrt{q}R|_{v}=\lim_{\mu\to0}\frac{R_{\square_v}}{\mu^3}$, which will be proven in next subsection. Notice that $\frac{p^i_{e_I}p^j_{e_J}}{V(v)}$ contains the inverse volume which complicates the expression. To avoid this problem, one can use Thiemann's trick
\begin{equation}
\epsilon_{jkl}\sqrt{q}e^{ak}e^{bl}=\epsilon^{abc}e_{cj}=2\epsilon^{abc}\{A_{cj}(x), V(x)\}/(\kappa\beta)
\end{equation}
to get another expression of $\tilde{R}$, where $V(x):=\int_{D }dy\sqrt{q}$ with $D\ni x$. We have
\begin{equation}
\tilde{R}=\int_{\Sigma} dx \sqrt{q}R(x)= -\frac{2}{\kappa\beta}\int_{\Sigma} dx R_{ab}^j\epsilon^{abc}\{A_{cj}(x), V(x)\}=\lim_{\mu\to0}\sum_{v\in \gamma_{\square}}R_{\square_v}
\end{equation}
 with
\begin{equation}\label{Rv2}
R_{\square_v}=-\frac{4}{\kappa\beta}\sum_{e_I,e_J,e_K\in \gamma_{\square}}^{e_I\cap e_J\cap e_K=v}\epsilon^{IJK}\text{tr} (h_{e_K}\{h^{-1}_{e_K}, V(v)\} h^{\Gamma}_{\square_{e_I,e_J}}),
\end{equation}
where $e_I,e_J, e_K$ have been re-oriented to be outgoing at $v$, $\epsilon^{IJK}=\text{sgn}[\det(e_I\wedge e_J\wedge e_K)]$, $\square_{e_I,e_J}$ is the minimal loop around a plaquette containing $e_I$ and $e_J$, which begins at $v$ via $e_I$ and gets back to $v$ through $e_J$.
\subsection{Geometric interpretation of $h_e^{\Gamma}$ and $R_{\square_v}$ from the continuum limit}
\subsubsection{Spin connection and scalar curvature in terms of triad}
Recall the expression of  the spin connection
\begin{equation}\label{gamma1}
\Gamma_{ajk}=-(\partial_a e_{bj}-\Gamma_{ab}^ce_{cj})e^b_k,
\end{equation}
which can also be expressed
 in terms of of $e_{a}^{i}$ as
\begin{equation}\label{gammaeee}
\Gamma_{a}^{i}=\frac{1}{2}\epsilon^{ijk}e_k^b
(\partial_be_{aj}-\partial_ae_{bj}+e^c_je_{al}\partial_be^l_c),
\end{equation}
with $\Gamma_{a}^{i}\epsilon_{ijk}\equiv\Gamma_{ajk}$.
Notice that the coordinate components of $\Gamma_{ab}^c$ transforms as a connection under the coordinate transformation. Let us introduce a regular coordinate system $\{x_I|I\in\{1,2,3\}\}|_p$ in a small open neighborhood of point $p$, which satisfies
 \begin{equation}\label{gf2}
\Gamma_{x_I x_J}^c|_{p}\equiv(\frac{\partial}{\partial x_I})^a(\frac{\partial}{\partial x_J})^b \Gamma_{ab}^c|_{p}=0, \quad \forall I\neq J.
 \end{equation}
This regular coordinate system $\{x_I|J\in\{1,2,3\}\}|_p$ can also be determined in a different way. Notice $\Gamma_{ajk}$ transforms as a connection under the gauge transformation.
Then, the regular coordinate system $\{x_I|J\in\{1,2,3\}\}|_p$ can be determined by requiring
  \begin{equation}\label{gf001}
\partial_{x_2} e_{x_1j}|_{p}=\partial_{x_2} e_{x_3j}|_{p}=\partial_{x_3} e_{x_1j}|_{p}=\partial_{x_3} e_{x_2j}|_{p}=\partial_{x_1} e_{x_2j}|_{p}=\partial_{x_1} e_{x_3j}|_{p}=0
 \end{equation}
 for the gauge choice
   \begin{equation}\label{gf000}
\Gamma_{ajk}|_{p}= 0.
 \end{equation}
  It is easy to verify that the regular coordinate systems $\{x_I|I\in\{1,2,3\}\}|_p$ determined by condition \eqref{gf2} or \eqref{gf001} are equivalent to each other, by using the definition of spin connection \eqref{gamma1}. Moreover, one should notice that once the regular coordinate system is determined by using the conditions \eqref{gf000} and \eqref{gf001}, the gauge fixing condition \eqref{gf000} can be released, since the gauge transformation of $\Gamma_{ajk}$ is independent with the coordinate transformation.

 Now, we can simplify the expression of $\Gamma_{a}^{i}|_p$ in this regular coordinate system $\{x_I|I\in\{1,2,3\}\}$ as follows. Denoted by $\Gamma_{x_Ijk}\equiv(\frac{\partial}{\partial x_I})^a\Gamma_{ajk}|$.
 Notice that $\Gamma_{ajk}$ transforms as a connection under the gauge transformation, and thus one can  make a gauge transformation which leads
 \begin{equation}\label{gf1}
\Gamma_{x_1jk}|_p\mapsto \Gamma_{x_1jk}|_p,\quad \Gamma_{x_2jk}|_p\mapsto 0,\quad \Gamma_{x_3jk}|_p\mapsto0.
 \end{equation}
Then, it is straightforwardly to get
 \begin{equation}\label{con1}
\partial_{x_2} e_{x_1j}|_p=\partial_{x_2} e_{x_3j}|_p=\partial_{x_3} e_{x_1j}|_p=\partial_{x_3} e_{x_2j}|_p=0
 \end{equation}
 by using
 \begin{equation}
\Gamma_{x_Ijk}|_p\equiv(\frac{\partial}{\partial x_I})^a\Gamma_{ajk}|_p=-(\partial_{x_I} e_{bj}-\Gamma_{x_Ib}^ce_{cj})e^b_k|_p,
\end{equation}
and the gauge condition \eqref{gf2} and \eqref{gf1}.
With the two conditions \eqref{gf2} and \eqref{gf1}, the expression of $\Gamma_{x_1}^i|_p$ can be simplified as
\begin{eqnarray}\label{gammax1}
\Gamma_{x_1}^{i}|_p&\equiv&(\frac{\partial}{\partial x_1})^a\Gamma_{a}^{i}|_p=\frac{1}{2}\epsilon^{ijk}e_k^b
(\partial_be_{x_1j}-\partial_{x_1}e_{bj}+e^c_je_{x_1l}\partial_be^l_c)|_p\\\nonumber
&=&\frac{1}{2}\epsilon^{ijk}
(-e_k^{x_2}\partial_{x_1}e_{x_2j}-e_k^{x_3}\partial_{x_1}e_{x_3j})|_p \\\nonumber &&+\frac{1}{2}\epsilon^{ijk}e_k^{x_1}(e^{x_2}_je_{x_1l}\partial_{x_1}e^l_{x_2})|_p +\frac{1}{2}\epsilon^{ijk}e_k^{x_1}(e^{x_3}_je_{x_1l}\partial_{x_1}e^l_{x_3})|_p
\end{eqnarray}
by using Eq.\eqref{con1}. Though the expression \eqref{gammax1} of $\Gamma_{x_1}^{i}|_p$ is given by considering the gauge transformation \eqref{gf1}, it is valid for arbitrary gauge choice since the gauge transformation \eqref{gf1} does not change $\Gamma_{x_1}^{i}|_p$. Similarly, one can give the expression of $\Gamma_{x_2}^{i}|_p\equiv{\frac{\partial}{\partial x_2}}^a\Gamma_{a}^{i}|_p$ and $\Gamma_{x_3}^{i}|_p\equiv{\frac{\partial}{\partial x_3}}^a\Gamma_{a}^{i}|_p$ in the regular coordinate system $\{x_I|I\in\{1,2,3\}\}$, which read,
\begin{eqnarray}\label{gammax2}
\Gamma_{x_2}^{i}|_p&\equiv&(\frac{\partial}{\partial x_2})^a\Gamma_{a}^{i}|_p=\frac{1}{2}\epsilon^{ijk}e_k^b
(\partial_be_{x_2j}-\partial_{x_2}e_{bj}+e^c_je_{x_2l}\partial_be^l_c)|_p\\\nonumber
&=&\frac{1}{2}\epsilon^{ijk}
(-e_k^{x_1}\partial_{x_2}e_{x_1j}-e_k^{x_3}\partial_{x_2}e_{x_3j})|_p \\\nonumber &&+\frac{1}{2}\epsilon^{ijk}e_k^{x_2}(e^{x_1}_je_{x_2l}\partial_{x_2}e^l_{x_1})|_p +\frac{1}{2}\epsilon^{ijk}e_k^{x_2}(e^{x_3}_je_{x_2l}\partial_{x_2}e^l_{x_3})|_p
\end{eqnarray}
and
\begin{eqnarray}\label{gammax3}
\Gamma_{x_3}^{i}|_p&\equiv&(\frac{\partial}{\partial x_3})^a\Gamma_{a}^{i}|_p=\frac{1}{2}\epsilon^{ijk}e_k^b
(\partial_be_{x_3j}-\partial_{x_3}e_{bj}+e^c_je_{x_3l}\partial_be^l_c)|_p\\\nonumber
&=&\frac{1}{2}\epsilon^{ijk}
(-e_k^{x_1}\partial_{x_3}e_{x_1j}-e_k^{x_2}\partial_{x_3}e_{x_2j})|_p \\\nonumber &&+\frac{1}{2}\epsilon^{ijk}e_k^{x_3}(e^{x_2}_je_{x_3l}\partial_{x_3}e^l_{x_2})|_p +\frac{1}{2}\epsilon^{ijk}e_k^{x_3}(e^{x_1}_je_{x_3l}\partial_{x_3}e^l_{x_1})|_p.
\end{eqnarray}

Let us now consider the scalar curvature $R$ determined by the spatial metric $q_{ab}$.
Using the equations \eqref{gammax1},\eqref{gammax2} and \eqref{gammax3}, one can try to express $R_{ab}^j|_p$ based on the triads and their derivatives in the regular coordinate system $\{x_I|I\in\{1,2,3\}\}|_p$. Note, however, that the derivative $\partial_{x_K}\Gamma_{x_L}^j|_p$ of the spin connection, which appears in the expression of $R_{x_Kx_L}^j$, contains the information of $\Gamma_{x_I}^j$ beyond the point $p$, which leads to Eqs. \eqref{gammax1},\eqref{gammax2} and \eqref{gammax3} for $\Gamma_{x_I}^j$ are not valid for the expression of $R_{x_Kx_L}^j$.

  To analyze the expression of the derivative $\partial_{x_K}\Gamma_{x_L}^j|_p$, consider the point $p^{\epsilon}_{K}$ whose coordinate satisfies $x_I(p^{\epsilon}_{K})=x_I(p)$ for $I\neq K$ and $x_K(p^{\epsilon}_{K})-x_K(p)=\epsilon$, where $\epsilon$ is small. Then, we have
  \begin{equation}\label{pgamma}
  \partial_{x_K}\Gamma_{x_L}^j|_p=\lim_{\epsilon\to 0}\frac{\Gamma_{x_L}^j|_{p^{\epsilon}_{K}}-\Gamma_{x_L}^j|_p}{\epsilon}.
  \end{equation}
  It should be noted that the expression of $\Gamma_{x_L}^j|_{p^{\epsilon}_{K}}$ is not directly given by Eqs. \eqref{gammax1}, \eqref{gammax2} and \eqref{gammax3} directly, since the regular coordinate system $\{x_I|I\in\{1,2,3\}\}|_p$ for $p$ may not be a regular coordinate system for $p^{\epsilon}_{K}$. Nevertheless,  one can always find a new coordinate system
   $\{\tilde{x}_I\}$
   which satisfies that, (i) $\tilde{x}_I(p^{\epsilon}_{K})=x_I(p^{\epsilon}_{K})$, $\tilde{x}_I(p)=x_I(p), \forall I$, and (ii) $\Gamma^{\tilde{x}_I}_{\tilde{x}_J\tilde{x}_L}|_p=\Gamma^{\tilde{x}_I}_{\tilde{x}_J\tilde{x}_L}|_{p^{\epsilon}_{K}}=0$, $\forall J\neq L$, where $\Gamma^{\tilde{x}_I}_{\tilde{x}_J\tilde{x}_L}\equiv \Gamma^a_{bc}(d\tilde{x}_I)_a(\frac{\partial}{\partial \tilde{x}_J})^b(\frac{\partial}{\partial \tilde{x}_L})^c$ and it is related to $\Gamma^{{x}_P}_{{x}_M{x}_N}$ by
    \begin{equation}
  \Gamma^{\tilde{x}_I}_{\tilde{x}_J\tilde{x}_L} =\sum_{M,N,P}\Gamma^{{x}_P}_{{x}_M{x}_N}\frac{\partial x_M}{\partial \tilde{x}_J}\frac{\partial x_N}{\partial \tilde{x}_L}\frac{\partial \tilde{x}_I}{\partial x_P}+\sum_{P}\frac{\partial^2 x_P}{\partial \tilde{x}_J\partial \tilde{x}_L}\frac{\partial \tilde{x}_I}{\partial x_P}.
  \end{equation}
  It is easy to see that the coordinate system $\{\tilde{x}_I\}$ satisfying above two conditions always exists, since the above two conditions for $\{\tilde{x}_I\}$ only involves specific two separate points.

Recalling the definition \eqref{gf2} of the  regular coordinate system, it is easy to see that $\{\tilde{x}_J|J\in\{1,2,3\}\}|_{p,p^{\epsilon}_{K}}$ is a regular coordinate system  for both $p$ and $p^{\epsilon}_{K}$. By taking $K=1,2,3$ step by step and establishing new coordinate system following the above procedures , we finally get a regular coordinate system $\{\tilde{\tilde{\tilde{x}}}_J|J\in\{1,2,3\}\}|_{p,p^{\epsilon}_1,p^{\epsilon}_2,p^{\epsilon}_3}$ for all of the points $p, p^{\epsilon}_1,p^{\epsilon}_2$ and $p^{\epsilon}_3$.
   To simplify our notations, we will still denote by $\{x_J|J\in\{1,2,3\}\}|_{p,p^{\epsilon}_1,p^{\epsilon}_2,p^{\epsilon}_3}$ the regular coordinate system for $p, p^{\epsilon}_1,p^{\epsilon}_2$ and $p^{\epsilon}_3$. Then, the expression of $\Gamma_{x_L}^j|_{p^{\epsilon}_{K}}, \forall K$ can be given by Eqs. \eqref{gammax1},\eqref{gammax2} and \eqref{gammax3} directly. Now, the scalar curvature $R$ at point $p$ can be expressed as
\begin{equation}\label{Rregular}
R|_{p}=-\sum_{K,L}R_{x_K x_L}^j\epsilon_{jmn}e^{x_Km}e^{x_Ln}|_{p}
\end{equation}
with
\begin{eqnarray}\label{Rxx}
R_{x_Kx_L}^j|_{p} &=& \partial_{x_K}\Gamma_{x_L}^j|_{p}-\partial_{x_L}\Gamma_{x_K}^j|_{p}+\epsilon^j_{\ mn}\Gamma_{x_K}^m\Gamma^n_{x_L}|_{p}\\
   &=&\lim_{\epsilon\to 0}\frac{\Gamma_{x_L}^j|_{p^{\epsilon}_{K}}-\Gamma_{x_L}^j|_p}{\epsilon}-\lim_{\epsilon\to 0}\frac{\Gamma_{x_K}^j|_{p^{\epsilon}_{L}}-\Gamma_{x_K}^j|_p}{\epsilon}+\epsilon^j_{\ mn}\Gamma_{x_K}^m\Gamma^n_{x_L}|_{p}
\end{eqnarray}
where we used Eq.\eqref{pgamma}, $\Gamma_{x_K}^j|_{p}$, $\Gamma_{x_L}^j|_{p}$ and $\Gamma_{x_K}^j|_{p^{\epsilon}_{L}}$, $\Gamma_{x_L}^j|_{p^{\epsilon}_{K}}$  are given by Eqs. \eqref{gammax1}, \eqref{gammax2} and \eqref{gammax3}. Notice that the expression \eqref{Rxx} of $R_{x_Kx_L}^j|_{p}$ relies on the regular coordinate system for both $p$ and the points $p^{\epsilon}_{K}$ near $p$. As we will seen in next subsection, such regular coordinate system appears in the shape-matching twisted geometry on cubic graph naturally, and it is crucial for the recovering of  the expression \eqref{Rregular} by the continuum limit of the discrete scalar curvature in the twisted geometry.

\subsubsection{Continuum limit of $h_e^{\Gamma}$ and $R_{\square_{v}}$}

\begin{figure}[h]
 \includegraphics[scale=0.16]{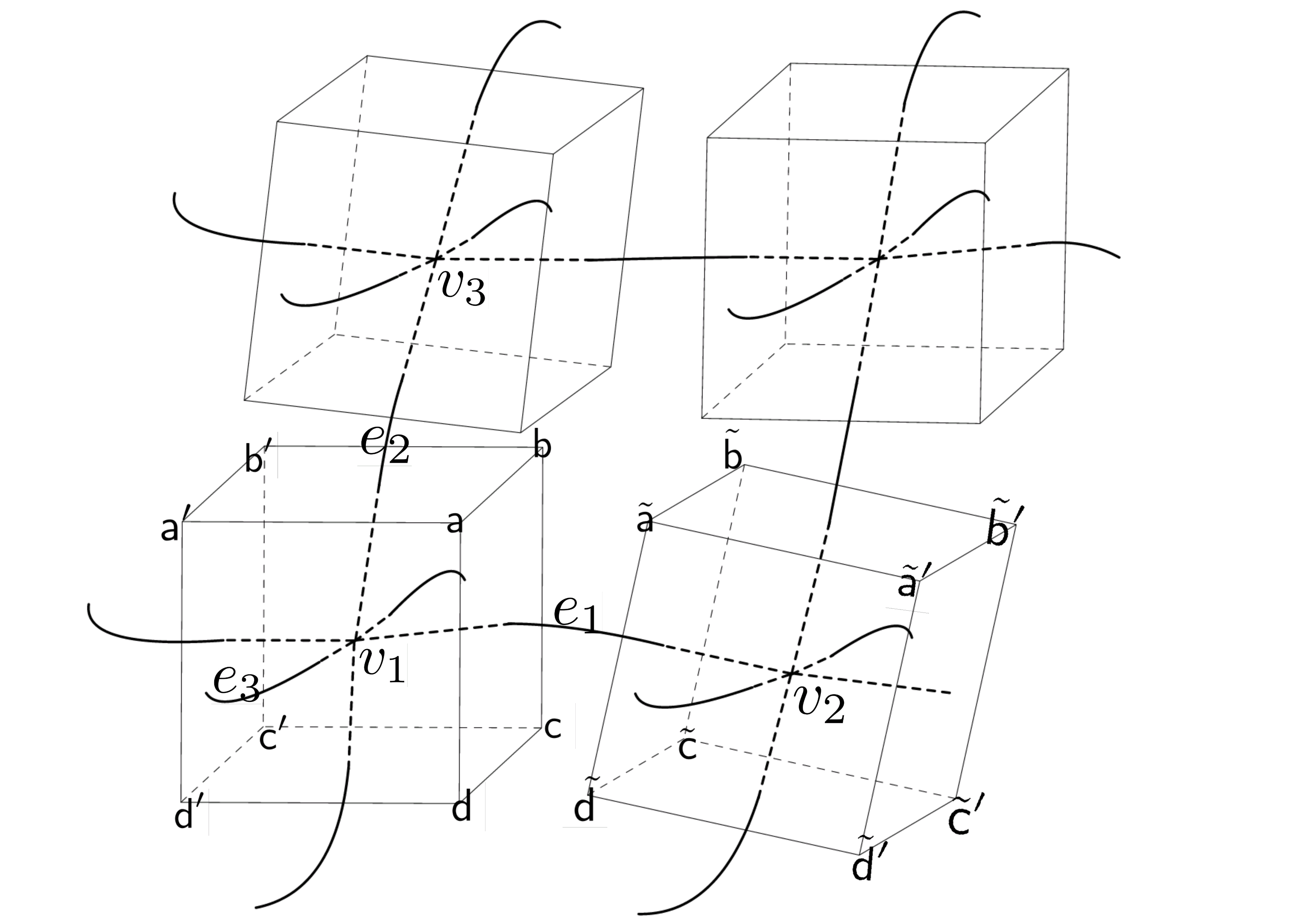}
\caption{A part of the cubic graph and the dual Polyhedras at each vertices. The edges in the cubic graph have the notations $e_1=e_1(v_1,+)=e_1(v_2,-)$, $e_2=e_2(v_1,-)=e_2(v_3,+)$ and $e_3=e_3(v_1,-)$.}
\label{fig2}
\end{figure}

Let us consider the continuum limit of the holonomy $h_e^{\Gamma}$ of spin connection on the edges in a cubic graph $\gamma_{\square}$, and focus on the configurations of twisted geometry which ensure that the shapes of the two faces dual to each edge are matched to each other.
We first adapt the cubic graph $\gamma_{\square}$ to a coordinate system $\{\hat{x}_I|I\in\{1,2,3\}\}$, with the coordinate length of each elementary  edge  ${e}_I(v,\pm)$ of $\gamma_{\square}$ being set to $\mu$ and the coordinate basis field satisfying $(\frac{\partial}{\partial \hat{x}_I})^a=\dot{e}_I^a(v,\pm)$ on each  ${e}_I(v,\pm)$, where the notation ${e}_I(v,\pm)$ ensures that $v$ is the source point of $e_I(v,+)$ and the target point of $e_I(v,-)$. For instance, one has $e_1=e_1(v_1,+)=e_1(v_2,-)$, $e_2=e_2(v_1,-)=e_2(v_3,+)$ and $e_3=e_3(v_1,-)$ in Fig. \ref{fig2}.

 Without loss of generality, our discussion will first focus on the single edge $e_1$ in figure \ref{fig2}. Then, we can introduce the following notations
\begin{equation}
w_{v_1,\hat{x}_1}^i\equiv \overrightarrow{\textsf{a}'\textsf{a}},\quad  w_{v_1,\hat{x}_2}^i\equiv \overrightarrow{\textsf{a}\textsf{d}},\quad w_{v_1,\hat{x}_3}^i\equiv \overrightarrow{\textsf{a}\textsf{b}}
\end{equation}
and
\begin{equation}
w_{v_2,\hat{x}_1}^i\equiv \overrightarrow{\tilde{\textsf{a}}\tilde{\textsf{a}}'},\quad  w_{v_2,\hat{x}_2}^i\equiv \overrightarrow{\tilde{\textsf{a}}\tilde{\textsf{d}}},\quad w_{v_2,\hat{x}_3}^i\equiv \overrightarrow{\tilde{\textsf{a}}\tilde{\textsf{b}}}
\end{equation}
 for the vectors in  figure \ref{fig2}. Correspondingly, one can define
 \begin{equation}
 w^{v_1,\hat{x}_1}_i:=\frac{p_{e_1,i}}{\text{V}(v_1)},\quad  w^{v_1,\hat{x}_2}_i:=-\frac{\tilde{p}_{e_2,i}}{\text{V}(v_1)},\quad w^{v_1,\hat{x}_3}_i:=-\frac{\tilde{p}_{e_3,i}}{\text{V}(v_1)}
 \end{equation}
 where $\square_{v_1}$ is the hexahedron dual to $v_1$,  and $\text{V}(v)$ is the volume of $\square_{v}$  defined by
\begin{equation}\label{Vdef}
\text{V}(v):=\sqrt{|\frac{1}{8}(\beta a^2 )^3\sum_{e\cap e'\cap e''=v}\epsilon_{ijk}\epsilon_{e,e',e''}p^i(v,e)p^j(v,e')p^k(v,e'')|}
\end{equation}
with $\epsilon_{e,e',e''}=\text{sgn}[\det(e\wedge e'\wedge e'')]$, $p^i(v,e)=p_e^i$ if $s(e)=v$ and $p^i(v,e)=-\tilde{p}_e^i$ if $t(e)=v$.
Based on these notations and recall the geometric interpretation of each factor of $h_e^{\Gamma}=n_e(V_e)e^{\zeta_e \tau_3}\tilde{n}_e(\tilde{V}_e)^{-1}$ shown in Fig.\ref{fig:label1}, one can check that the generator of $h_{e_1}^{\Gamma}=\exp(\Gamma_{e_1}^{i}\tau_i)$ is given by
\begin{eqnarray}\label{gammax1ge}
\Gamma_{e_1}^{i}&=&\frac{1}{2}\epsilon^{ijk}
(-w_k^{v_1,\hat{x}_2}{(w_{v_2,\hat{x}_2j}-w_{v_1,\hat{x}_2j})}-w_k^{v_1,\hat{x}_3}{(w_{v_2,\hat{x}_3j} -w_{v_1,\hat{x}_3j})})(1+\mathcal{O}(\mu)) \\\nonumber &&+\frac{1}{2}\epsilon^{ijk}w_k^{v_1,\hat{x}_1}(w^{v_1,\hat{x}_2}_jw_{v_1,\hat{x}_1l}{(w^l_{v_2,\hat{x}_2} -w^l_{v_1,\hat{x}_2})}) (1+\mathcal{O}'(\mu)) \\\nonumber &&+\frac{1}{2}\epsilon^{ijk}w_k^{v_1,\hat{x}_1}(w^{v_1,\hat{x}_3}_jw_{v_1,\hat{x}_1l} {(w^l_{v_2,\hat{x}_3}-w^l_{v_1,\hat{x}_3})})(1+\mathcal{O}''(\mu))
\end{eqnarray}
for small $|\Gamma_{e_1}^{i}| \propto \mu$ with $\mu$ the lattice length scale. We leave the detailed check in Appendix \ref{app:check_continuum}.
Further, one can check that the spin connection along $e_1$ can be given by
\begin{eqnarray}\label{gammax1mu}
\Gamma_{\hat{x}_1}^{i}|_{v_1}&=&\lim_{\mu\to0} \frac{\Gamma_{e_1}^{i}}{\mu}\\\nonumber
&=&\frac{1}{2}\epsilon^{ijk}
(-e_k^{\hat{x}_2}\partial_{\hat{x}_1}e_{\hat{x}_2j}-e_k^{\hat{x}_3}\partial_{\hat{x}_1}e_{\hat{x}_3j})|_{v_1} \\\nonumber &&+\frac{1}{2}\epsilon^{ijk}e_k^{\hat{x}_1}(e^{\hat{x}_2}_je_{\hat{x}_1l}\partial_{\hat{x}_1}e^l_{\hat{x}_2})|_{v_1} +\frac{1}{2}\epsilon^{ijk}e_k^{\hat{x}_1}(e^{\hat{x}_3}_je_{\hat{x}_1l}\partial_{\hat{x}_1}e^l_{\hat{x}_3})|_{v_1}
\end{eqnarray}
by introducing the continuum limit
\begin{equation}
e^{\hat{x}_I}_j|_{v_1}=\lim_{\mu\to0}(\mu w^{v_1,\hat{x}_I}_j),\quad  e_{\hat{x}_I}^j|_{v_1}=\lim_{\mu\to0}\frac{ w_{v_1,\hat{x}_I}^j}{\mu}
\end{equation}
and
\begin{equation}
\partial_{\hat{x}_1}e_{\hat{x}_2}^i|_{v_1}=\lim_{\mu\to0}\frac{(w_{v_2,\hat{x}_2}^i-w_{v_1,\hat{x}_2}^i)}{\mu^2},\quad \partial_{\hat{x}_1}e_{\hat{x}_3}^i|_{v_1}=\lim_{\mu\to0}\frac{(w_{v_2,\hat{x}_3}^i-w_{v_1,\hat{x}_3}^i)}{\mu^2}.
\end{equation}
Finally, by comparing Eqs.\eqref{gammax1} and \eqref{gammax1mu}, one can conclude that the continuum limit of the generator $\Gamma_{e_1}^{i}$ of $h_e^{\Gamma}$ reproduces the spin connection defined by triad exactly for the cubic graph.

 Nevertheless, it should be noted that Eq.\eqref{gammax1} gives the expression of $\Gamma_{x_1}^i$ at $p$ based on the regular coordinate system $\{x_I\}$. Thus, one must require that $\{\hat{x}_I\}$ is also a regular coordinate system in the continuum limit to ensure the consistency of eqs.\eqref{gammax1} and \eqref{gammax1mu}.
In fact, for the Regge geometry sector of  twisted geometry which satisfies the shape-matching condition, one can immediately get that $\Gamma_{e_1}^{i}=0$ leads to $\Gamma_{\hat{x}_1}^{i}|_{v_1}=0$ and $\partial_{\hat{x}_1}e_{\hat{x}_2}^i=\partial_{\hat{x}_1}e_{\hat{x}_3}^i$ in the limit $\mu\to0$, and this result can be generalized to $\Gamma_{e_2}^{i}=0$ and $\Gamma_{e_3}^{i}=0$ directly. Thus, the coordinate system $\{\hat{x}_I\}$ considered here is actually a regular coordinate system in the continuum limit for the shape-matched twisted geometry.
Finally, one can conclude that the continuum limit of $h_e^{\Gamma}$ reproduces the spin connection defined by triad exactly for the cubic graph for the shape-matching twisted geometry. 

{ It is also worth to have an analysis on the continuum limit of $h_{e_1}^{\Gamma}$ in the non-Regge geometry sector of twisted geometry. Recall the dependence of $h_{e_1}^{\Gamma}=h_{e_1,\square}^{\Gamma}$ on a minimal loop $\square$ containing ${e_1}$, one has the collection $\{\square_1,\square_2,\square_3,\square_4\}$ of minimal loops, with  $e_1,e_2(v_1,+)\subset\square_1 $, $e_1,e_2(v_1,-)\subset\square_2 $ , $e_1,e_3(v_1,+)\subset\square_3 $ and $e_1,e_3(v_1,-)\subset\square_4 $ as shown in Fig.\ref{fig2}. Then, it is straightforwardly to give the continuum limit of $h_{e_1}^{\Gamma}=h_{e_1, \square}^{\Gamma}$ , which reads
\begin{eqnarray}\label{gammax1munonregge}
\lim_{p\stackrel{e_2(v_1,+)}{\longrightarrow} v_1}\Gamma_{\hat{x}_1}^{i}(p)&=&\lim_{\mu\to0} \frac{\Gamma_{e_1,\square_1}^{i}}{\mu},\quad \lim_{p\stackrel{e_2(v_1,-)}{\longrightarrow} v_1}\Gamma_{\hat{x}_1}^{i}(p)=\lim_{\mu\to0} \frac{\Gamma_{e_1,\square_2}^{i}}{\mu},\\\nonumber
\lim_{p\stackrel{e_3(v_1,+)}{\longrightarrow} v_1}\Gamma_{\hat{x}_1}^{i}(p)&=&\lim_{\mu\to0} \frac{\Gamma_{e_1,\square_3}^{i}}{\mu},\quad \lim_{p\stackrel{e_3(v_1,-)}{\longrightarrow} v_1}\Gamma_{\hat{x}_1}^{i}(p)=\lim_{\mu\to0} \frac{\Gamma_{e_1,\square_4}^{i}}{\mu},
\end{eqnarray}
where $p\stackrel{e_2(v_1,+)}{\longrightarrow} v_1$ represents that the limit is taken from $p$ to $v_1$ along the path $e_2(v_1,+)$ and likewise for $e_2(v_1,-)$, $e_3(v_1,+)$ and $e_3(v_1,-)$,  $\Gamma_{e_1,\square_1}^{i}$ is the generator of $h_{e_1,\square_1}^{\Gamma}=\exp(\Gamma_{e_1,\square_1}^{i}\tau_i)$  and likewise for $\Gamma_{e_1,\square_2}^{i}$, $\Gamma_{e_1,\square_3}^{i}$ and $\Gamma_{e_1,\square_4}^{i}$. Indeed, this result shows that the continuum limit of $h_{e_1}^{\Gamma}$ in the non-Regge geometry sector of twisted geometry gives a distributional spin-connection at $v_1$.
Let us have a further discussion on the continuum limit of  $h_e^{\Gamma}$ in the non-Regge geometry sector of  twisted geometry. Recall that both of the classical expression\eqref{gammaeee} for the spin-connection  and  \eqref{gf001}  for  the regular coordinate involve the derivative of $e_a^i$, which means these definitions only hold for the triad field which is differentiable at every point of the spatial manifold. Nevertheless, one should notice that, in the procedure of the construction of quantum configuration space of LQG \cite{thiemann2008modern,Ashtekar:2004eh,rovelli_vidotto_2014,Han2005FUNDAMENTAL}, the loop representation  extends the consideration to the general distributional field, for which the expressions  \eqref{gammaeee} and  \eqref{gf001} defined in classical theory  is meaningless. In fact, the general distributional field introduced by the loop representation is purely quantum degrees of freedom, hence it has  definite geometric interpretation only in discrete theory. Then, let us go back to the  the continuum limit of non-Regge geometry sector of  twisted geometry.  Notice that the regular coordinate determined by  \eqref{gf001} always exist for the classical and differentiable triad field. However, for the non-Regge geometry sector of  twisted geometry which does not satisfy the shape-matching condition, it is not able to establish a coordinate $\{\hat{x}_i\}$ in Fig.\ref{fig2} to  give the relation $\partial_{\hat{x}_1}e_{\hat{x}_2}^i=\partial_{\hat{x}_1}e_{\hat{x}_3}^i$, so that it is impossible to give the regular coordinate  in the limit $\mu\to0$. Hence, the continuum limit of the non-Regge geometry sector of  twisted geometry does not give the classical and differentiable triad field, but it gives the generalized distributional triad field which only has  definite geometric interpretation  in discrete theory. In fact, similar analysis has been mentioned in Ref.\cite{PhysRevD.87.024038}, in which the spin connection for a couple of glued tetrahedron in twisted geometry is established as a distribution field, and its continuum limit recovers the classical connection of triad field only for the Regge geometry sector.}

\begin{figure}[h]
 \includegraphics[scale=0.16]{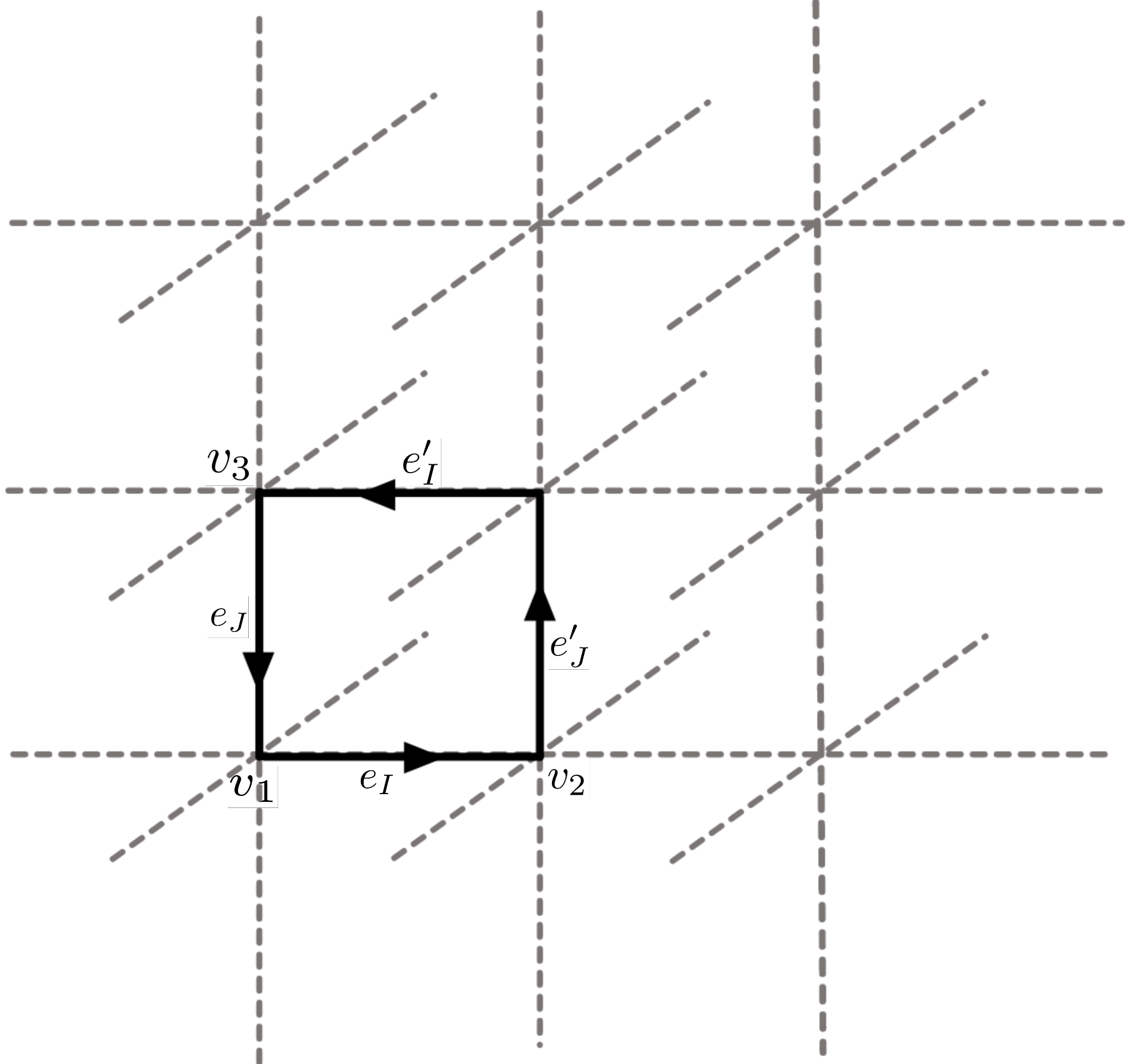}
\caption{The minimal loop $\square_{e_I,e_J}=e_I\circ e'_J\circ e'_I\circ e_J$ containing edges $e_I$ and $e_J$, which begins at $v_1$ via $e_I$ and gets back to $v_1$ through $e_J$.}
\label{fig3}
\end{figure}

It is now ready to consider the continuum limit of $R_{\square_{v_1}}$ based on the continuum limit of $h_e^\Gamma$.  Recall that the expression \eqref{Rv1} of $R_{\square_{v_1}}$ is given by
\begin{equation}
R_{\square_{v_1}}=\frac{(\beta a^2)^2}{2}\sum_{\square_{e_I,e_J}}\frac{p^i_{e_I}p^j_{e_J}}{V(v_1)}\epsilon_{ijk}\text{tr} (\tau^kh^{\Gamma}_{\square_{e_I,e_J}})
\end{equation}
on the cubic graph, where $\square_{e_I,e_J}=e_I\circ e'_{J}\circ e'_{I}\circ e_J$, $e_I=e_I(v_1,+)=e_I(v_2,-),  e'_{J}=(e_J(v_2,-))^{-1}, e'_{I}=(e_I(v_3,+))^{-1},  e_J=e_J(v_1,-)=e_J(v_3,+)$ as shown in Fig. \ref{fig3}, and $h^{\Gamma}_{\square_{e_I,e_J}}=h^{\Gamma}_{e_I}h^{\Gamma}_{e'_{J}}h^{\Gamma}_{e'_{I}}h^{\Gamma}_{e_J}$. By using $h^{\Gamma}_{e_I}=\exp(\mu \Gamma_{e_I}^i\tau_i)$, one can get
\begin{equation}
-2\text{tr} (\tau^kh^{\Gamma}_{\square_{e_I,e_J}})=\mu^2(\frac{\Gamma^k_{e'_I} +\Gamma^k_{e_I}}{\mu}+ \frac{\Gamma^k_{e'_J} +\Gamma^k_{e_J}}{\mu}-\epsilon^k_{\ mn}\Gamma^m_{e_I}\Gamma^n_{e_J})+\mathcal{O}({\mu^3}).
\end{equation}
Thus, we have
\begin{eqnarray}
 &&  \lim_{\mu\to0}\frac{-2\text{tr} (\tau^kh^{\Gamma}_{\square_{e_I,e_J}})}{\mu^2}\\\nonumber
   &=& \lim_{\mu\to0} \frac{\Gamma^k_a\dot{e}'^a_J|_{v_2} +\Gamma^k_b\dot{e}^b_J|_{v_1}}{\mu} + \lim_{\mu\to0}\frac{\Gamma^k_a\dot{e}'^a_I |_{v_3} +\Gamma^k_b\dot{e}^b_I|_v}{\mu}-\epsilon^k_{\ mn}\Gamma^m_a\dot{e}^a_I\Gamma^n_b\dot{e}^b_J|_{v_1},
\end{eqnarray}
where $v_2$ is the target point of $e_I$ and $v_3$ is the source point of $e_J$ as shown in Fig.\ref{fig3}.
Furthermore, by using Eq.\eqref{gammax1mu}, one get
\begin{eqnarray}
 \lim_{\mu\to0}\frac{-2\text{tr} (\tau^kh^{\Gamma}_{\square_{e_I,e_J}})}{\mu^2}&=&  \lim_{\mu\to0} \frac{\Gamma^k_{\hat{x}_J}|_{v_2} -\Gamma^k_{\hat{x}_J}|_{v_1}}{\mu} - \lim_{\mu\to0}\frac{\Gamma^k_{\hat{x}_I} |_{v_3} -\Gamma^k_{\hat{x}_I}|_{v_1}}{\mu}+\epsilon^k_{\ mn}\Gamma^m_{\hat{x}_I}\Gamma^n_{\hat{x}_J}|_{v_1}\\\nonumber
   &=&  \partial_{\hat{x}_I}\Gamma^k_{\hat{x}_J}|_{v_1} -\partial_{\hat{x}_J}\Gamma^k_{\hat{x}_I}|_{v_1}+\epsilon^k_{\ mn}\Gamma^m_{\hat{x}_I}\Gamma^n_{\hat{x}_J}|_{v_1}\\\nonumber
   &=&R_{\hat{x}_I\hat{x}_J}^k|_{v_1},
\end{eqnarray}
where we use Eq.\eqref{Rxx} in the third ``$=$'' and the fact $R_{\hat{x}_I\hat{x}_J}^k|_{v_1}:=R_{ab}^k\dot{e}_I^a\dot{e}_J^b|_{v_1}$. Moreover, the continuum limit of the remaining factor $\frac{p^i_{e_I}p^j_{e_J}}{V(v_1)}\epsilon_{ijk}$ in $R_{\square_{v_1}}$ is given by
\begin{eqnarray}
 \lim_{\mu\to0}\frac{(\beta a^2)^2}{\mu}\frac{p^i_{e_I}p^j_{e_J}}{V(v_1)}\epsilon_{ijk}&=& |\det{(e)}| e^{\hat{x}_I,i}e^{\hat{x}_J,j}\epsilon_{ijk}|_{v_1},
\end{eqnarray}
where we used that $\lim_{\mu\to0}\frac{V_{v_1}}{\mu^3}=|\det(e)|_{v_1}$ and $\lim_{\mu\to0}\frac{\beta a^2p^i_{e_I}}{\mu^2}=|\det{(e)}|\cdot e^{\hat{x}_I,i}| _{v_1}$ with $e^{\hat{x}_I,i}| _{v_1}:=e^{ai}(d\hat{x}_I)_a| _{v_1}$. Finally, we have the limit
\begin{equation}
\lim_{\mu\to0}\frac{R_{\square_{v_1}}}{\mu^3}=-|\det(e)|\sum_{I,J} e^{\hat{x}_I,i}e^{\hat{x}_J,j}R_{\hat{x}_I\hat{x}_J}^k\epsilon_{ijk}|_{v_1}
\end{equation}
which is the expression of the densitized scalar curvature $|\det(e)| R$ at $v_1$ in the regular coordinate system $\{\hat{x}_I\}$ exactly. Hence, we conclude that the $R_{\square_{v_1}}$ gives correct continuum limit for $\mu\to 0$. Following a similar procedures, one can get the same result for the expression \eqref{Rv2} of $R_{\square_{v_1}}$.
\section{Quantization of the holonomy of spin connection and spatial scalar curvature}\label{sec4}
The holonomy $h^{\Gamma}$ of the spin connection has been expressed in terms of fluxes by Eqs.\eqref{hgammafinal}.
Its quantization can be given by directly replacing the fluxes with the corresponding flux operators. This can be done step by step as follows. First, let us introduce the notations
\begin{equation}
\Big(\sqrt{\hat{Y}}\Big)^{-1}:=\sum_{Y\in\mathcal{E}\setminus 0}\sqrt{Y}^{-1}|Y\rangle\langle Y|,
\end{equation}
\begin{equation}
\sqrt{\hat{Y}}:=\sum_{Y\in\mathcal{E}}\sqrt{Y}|Y\rangle\langle Y|
\end{equation}
and
\begin{equation}
\text{sgn}(\hat{Y}):=\sum_{Y\in\mathcal{E}}\text{sgn}(Y)|Y\rangle\langle Y|,
\end{equation}
where $\mathcal{E}$ is the eigen-spectrum of $\hat{Y}$ and $|Y\rangle$ is the eigen-state which corresponding to the eigenvalue $Y$ of $\hat{Y}$ \cite{Bianchi:2008es,Ma:2010fy}.
Then,  $V^i_e=\frac{p_e^i}{\sqrt{p_e^ip_{e,i}}}$ and $\tilde{V}^i_e=\frac{\tilde{p}_e^i}{\sqrt{\tilde{p}_e^i\tilde{p}_{e,i}}}$ can be promoted as the operators
\begin{equation}\label{hatV}
 \hat{ V}^i_e:=\hat{p}_e^i\Big(\sqrt{\hat{p}_e^i\hat{p}_{e,i}}\Big)^{-1}
\end{equation}
and
\begin{equation}\label{hatVt}
\hat{\tilde{V}}^i_e:=\hat{\tilde{p}}_e^i\Big(\sqrt{\hat{\tilde{p}}_e^i\hat{\tilde{p}}_{e,i}}\Big)^{-1}
\end{equation}
respectively.
Furthermore, based on Eqs.\eqref{hatV} and \eqref{hatVt}, one can  define the operators
\begin{eqnarray}
\widehat{w_0(\Gamma_e)}
&:=&- \frac{\widehat{\sin(\zeta_e/2)}}{2}\Big(\sqrt{1+\hat{V}_e^3}\Big)^{-1} \Big(\sqrt{1-\hat{\tilde{V}}_e^3}\Big) ^{-1} (\hat{V}_e^2\hat{\tilde{V}}_e^1 -\hat{V}_e^1\hat{\tilde{V}}_e^2)\\\nonumber
&&+\frac{\widehat{\cos(\zeta_e/2)}}{2}\Bigg(\sqrt{1-\hat{\tilde{V}}_e^3}\sqrt{1+\hat{V}_e^3}-\Big(\sqrt{1+\hat{V}_e^3}\Big)^{-1} \Big(\sqrt{1-\hat{\tilde{V}}_e^3}\Big)^{-1} (\hat{V}_e^1\hat{\tilde{V}}_e^1+\hat{V}_e^2\hat{\tilde{V}}_e^2)\Bigg),
\end{eqnarray}
\begin{eqnarray}
\widehat{w_1(\Gamma_e)}
&:=&\frac{\widehat{\cos(\zeta_e/2)}}{2}\Bigg(\sqrt{1-\hat{\tilde{V}}_e^3}\Big(\sqrt{1+\hat{V}_e^3}\Big)^{-1}\hat{V}_e^2 +\sqrt{1+\hat{V}_e^3} \Big(\sqrt{1-\hat{\tilde{V}}_e^3}\Big)^{-1}\hat{\tilde{V}}_e^2\Bigg) \\\nonumber
&& +\frac{\widehat{\sin(\zeta_e/2)}}{2}\Bigg(\sqrt{1+\hat{V}_e^3} \Big(\sqrt{1-\hat{\tilde{V}}_e^3}\Big)^{-1}  \hat{\tilde{V}}_e^1-\sqrt{1-\hat{\tilde{V}}_e^3}\Big(\sqrt{1+\hat{V}_e^3}\Big)^{-1} \hat{V}_e^1\Bigg),
\end{eqnarray}
\begin{eqnarray}
\widehat{w_2(\Gamma_e)}&:=&-\frac{\widehat{\cos(\zeta_e/2)}}{2}\Bigg(\sqrt{1+\hat{V}_e^3} \Big(\sqrt{1-\hat{\tilde{V}}_e^3}\Big)^{-1}\hat{\tilde{V}}_e^1 +\sqrt{1-\hat{\tilde{V}}_e^3}\Big(\sqrt{1+\hat{V}_e^3}\Big)^{-1}\hat{V}_e^1 \Bigg)\\\nonumber
&&
  +\frac{\widehat{\sin(\zeta_e/2)}}{2}\Bigg(\sqrt{1+\hat{V}_e^3} \Big(\sqrt{1-\hat{\tilde{V}}_e^3}\Big)^{-1}  \hat{\tilde{V}}_e^2-\sqrt{1-\hat{\tilde{V}}_e^3}\Big(\sqrt{1+\hat{V}_e^3}\Big)^{-1}\hat{ V}_e^2\Bigg)
\end{eqnarray}
and
\begin{eqnarray}
\widehat{w_3(\Gamma_e) }&:=& -\frac{\widehat{\cos(\zeta_e/2)}}{2}\Big(\sqrt{1+\hat{V}_e^3}\Big)^{-1} \Big(\sqrt{1-\hat{\tilde{V}}_e^3}\Big)^{-1} (-\hat{V}_e^2\hat{\tilde{V}}_e^1 +\hat{V}_e^1\hat{\tilde{V}}_e^2) \\\nonumber
&& +\frac{ \widehat{\sin(\zeta_e/2)}}{2}\Big(-\sqrt{1-\hat{\tilde{V}}_e^3}\sqrt{1+\hat{V}_e^3}+\Big(\sqrt{1+\hat{V}_e^3}\Big)^{-1} \Big(\sqrt{1-\hat{\tilde{V}}_e^3}\Big)^{-1} (-\hat{V}_e^1\hat{\tilde{V}}_e^1 -\hat{V}_e^2\hat{\tilde{V}}_e^2) \Big),
\end{eqnarray}
with $\widehat{\sin(\zeta_e/2)}$ and $\widehat{\cos(\zeta_e/2)}$ being respectively defined by
\begin{equation}
\widehat{\sin(\zeta_e/2)}:=\sqrt{\frac{1+\widehat{\cos\zeta_e}}{2}}
\end{equation}
and
\begin{equation}
\widehat{\cos(\zeta_e/2)}:=\widehat{\text{sgn}(\zeta_e)}\sqrt{\frac{1-\widehat{\cos\zeta_e}}{2}},
\end{equation}
where $\widehat{\cos\zeta_e}$ and $\widehat{\text{sgn}(\zeta_e)}$ are given by
\begin{eqnarray}
\widehat{\cos\zeta_e}& =&\Big(\widehat{\bar{V}}_{e_\imath}^i\widehat{\bar{V}}_{\tilde{e}_\imath}^j\delta_{ij} -\hat{V}_{e_\imath}^k\hat{V}_{e,k}\widehat{\bar{V}}_{\tilde{e}_\imath}^3 + \hat{V}_{\tilde{e}_\imath}^k\hat{\tilde{V}}_{e,k}\widehat{\bar{V}}_{e_\imath}^3 -\hat{V}_{\tilde{e}_\imath}^k\hat{\tilde{V}}_{e,k}\hat{V}_{e_\imath}^l\hat{V}_{e,l}\Big) \\\nonumber &&\cdot\Big(\sqrt{1-(\hat{V}_{e_\imath}^k\hat{V}_{e,k})^2}\Big)^{-1} \Big(\sqrt{1-(\hat{V}_{\tilde{e}_\imath}^k\hat{\tilde{V}}_{e,k})^2}\Big)^{-1}
\end{eqnarray}
and
\begin{equation}
\widehat{\text{sgn}(\zeta_e)} =\text{sgn}(\epsilon_{ijk}\delta_3^i\widehat{\bar{V}}_{\tilde{e}_{\imath}}^j\widehat{\bar{V}}_{e_\imath}^k)
\end{equation}
with
\begin{eqnarray}
\widehat{\bar{V}}_{\tilde{e}_\imath}^i
&=&\delta_3^j\hat{\tilde{V}}_e^k{\epsilon_{jk}}^i(\hat{\tilde{V}}_e^{1} \hat{V}_{\tilde{e}_{\imath}}^{2} -\hat{\tilde{V}}_e^{2} \hat{V}_{\tilde{e}_{\imath}}^{1} )\Big(1-(\hat{\tilde{V}}_e^3)^2\Big)^{-1}\\\nonumber
&&+\Big(\big(2(\hat{\tilde{V}}_e^{3})^2-1\big)\delta_3^{i} \hat{\tilde{V}}_e^{j}\hat{V}_{\tilde{e}_{\imath},j}-\hat{\tilde{V}}_e^{3}\delta_3^{i}\hat{\tilde{V}}_e^{k} \hat{\tilde{V}}_{e,k}\hat{V}_{\tilde{e}_{\imath},3} +\hat{\tilde{V}}_e^{i} \hat{V}_{\tilde{e}_{\imath}}^3-\hat{\tilde{V}}_e^{i}\hat{\tilde{V}}_e^{3} \hat{\tilde{V}}_e^{j}\hat{V}_{\tilde{e}_{\imath},j}\Big)\Big(1-(\hat{\tilde{V}}_e^3)^2\Big)^{-1}
\end{eqnarray}
and
\begin{eqnarray}
\widehat{\bar{V}}_{e_\imath}^i
&=&\delta_3^j\hat{V}_e^k{\epsilon_{jk}}^i(\hat{V}_e^{1}\hat{V}_{{e}_{\imath}}^{2}-\hat{V}_e^{2} \hat{V}_{{e}_{\imath}}^{1})\Big(1-(\hat{V}_e^3)^2\Big)^{-1}\\\nonumber
 &&+\Big(\big(1-2(\hat{V}_e^3)^2\big)\delta_3^{i} \hat{V}_e^{j}\hat{V}_{{e}_{\imath},j} +\hat{V}_e^3\delta_3^{i}\hat{V}_{e,k'} \hat{V}_e^{k'}\hat{V}^3_{{e}_{\imath}}+\hat{V}_e^3\hat{V}_e^{i} \hat{V}_e^{j}\hat{V}_{{e}_{\imath},j}-\hat{V}_e^{i} \hat{V}^3_{{e}_{\imath}}\Big)\Big(1-(\hat{V}_e^3)^2\Big)^{-1}.
\end{eqnarray}
Now, recall $h_e^{\Gamma}:=w_0(\Gamma_e)\mathbb{I}+w_1(\Gamma_e)\mathbf{i}\sigma_1+w_2(\Gamma_e)\mathbf{i}\sigma_2 +w_3(\Gamma_e)\mathbf{i}\sigma_3$ and notice that $w_0(\Gamma_e)$, $w_1(\Gamma_e)$, $w_2(\Gamma_e)$ and $ w_3(\Gamma_e)$ are real, the holonomy operator of spin connection can be given by
\begin{equation}\label{hnnop}
\widehat{h_e^{\Gamma}}=\frac{\widehat{w_0(\Gamma_e)}+\widehat{w_0(\Gamma_e)}^\dagger}{2}\mathbb{I} +\frac{\widehat{w_1(\Gamma_e)}+\widehat{w_1(\Gamma_e)}^\dagger}{2}\mathbf{i}\sigma_1 +\frac{\widehat{w_2(\Gamma_e)}+\widehat{w_2(\Gamma_e)}^\dagger}{2}\mathbf{i}\sigma_2 +\frac{\widehat{w_3(\Gamma_e)}+\widehat{w_3(\Gamma_e)}^\dagger}{2}\mathbf{i}\sigma_3.
\end{equation}

The quantization of $\tilde{R}$ can be  given by substituting the fluxes in \eqref{Rv1} with the corresponding flux operators directly. We define
 \begin{equation}
\widehat{h^{\Gamma}}_{\square_{e_I,e_J}} :=\widehat{h^{\Gamma}_{e_I}}\widehat{h^{\Gamma}_{e'_{J}}}\widehat{h^{\Gamma}_{e'_{I}}} \widehat{h^{\Gamma}_{e_J}},
 \end{equation}
 where $\widehat{h^{\Gamma}_{e_I}}, \widehat{h^{\Gamma}_{e'_{J}}}, \widehat{h^{\Gamma}_{e'_{I}}}, \widehat{h^{\Gamma}_{e_J}}$ are defined by Eq.\eqref{hnnop} based on the minimal loop $_{\square_{e_I,e_J}}$.
Then, we can  define the spatial curvature operator on cubic graph $\gamma_\square$ as
\begin{equation}\label{sumRvop}
\hat{\tilde{R}}:=\sum_{v\in \gamma_{\square}}\hat{R}_{\square_v},
\end{equation}
where $\hat{R}_{\square_v}$ is defined by $\hat{R}_{\square_v}:=\frac{1}{2}(\hat{\mathfrak{R}}_{\square_v}+\hat{\mathfrak{R}}^\dagger_{\square_v})$ with
\begin{equation}
\hat{\mathfrak{R}}_{\square_v}:=\frac{(\beta a^2)^2}{2}\sum_{\square_{e_I,e_J}}\hat{V}(v)^{-1}\hat{p}^i_{e_I}\hat{p}^j_{e_J}\epsilon_{ijk}\text{tr} (\tau^k\widehat{h^{\Gamma}}_{\square_{e_I,e_J}}).
\end{equation}
Similarly, one can also define  $\hat{R}_{\square_v}$ based on another regularization expression \eqref{Rv2}, in which the inverse volume operator is avoided by using e Thiemann's trick. The corresponding result reads $\hat{R}_{\square_v}:=\frac{1}{2}(\hat{\mathfrak{R}}_{\square_v}+\hat{\mathfrak{R}}^\dagger_{\square_v})$ with
\begin{equation}
\hat{\mathfrak{R}}_{\square_v}=-\frac{4}{\mathbf{i}\hbar\kappa\beta}\sum_{e_I,e_J,e_K\in \gamma_{\square}}^{e_I\cap e_J\cap e_K=v}\epsilon^{IJK}\text{tr} (h_{e_K}[h^{-1}_{e_K}, \hat{V}(v)]\widehat{ h^{\Gamma}}_{\square_{e_I,e_J}}).
\end{equation}
Note that $\hat{\mathfrak{R}}_{\square_v}$ is composed purely of flux and holonomy operators, so it and its adjoint $\hat{\mathfrak{R}}^\dagger_{\square_v}$ are well-defined in the smooth cylindrical function space $\text{Cyl}_{\gamma_\square}^{\infty}$ on $\gamma_{\square}$.

\section{Conclusion and outlook}\label{sec5}
  A new construction of the spatial scalar curvature operator in (1+3)-dimensional LQG is proposed in this article. Since the previous constructions encounter certain problems, a new strategy for the construction of the spatial scalar curvature operator based on the twisted geometry is considered. More explicitly, the holonomy of the spin connection is expressed in terms of the twisted geometry variables, and it is checked that its generator recovers the spin connection  in the case of shape-matching in a certain continuum limit.
  The spatial scalar curvature in terms of the twisted geometry is given by the composition of the holonomy of the spin connection on the loops. Finally, by using the twisted geometry parametrization of the holonomy-flux phase space and replacing the fluxes with flux operators, the holonomy of the spin connection and the spatial scalar curvature in terms of twisted geometry variables are promoted as well-defined operators.

A few points are worth discussing. {First, since there is no shape-matching condition for twisted geometry, the establishment of holonomy operator $\widehat{h^{\Gamma}_{e}}$ of the spin connection  on $e$ relies on a minimal loop containing $e$, and the choice of such a minimal loop introduces an ambiguity for the construction of $\widehat{h^{\Gamma}_{e}}$. Nevertheless, $\widehat{h^{\Gamma}_{e}}$ always associates a minimal loop when it appears in the spatial scalar curvature operator $\hat{\tilde{R}}$, so this ambiguity automatically disappears in the construction of $\hat{\tilde{R}}$. Especially, this is also an advantage of our construction for the holonomy of the spin connection compared to the results in Ref. \cite{PhysRevD.87.024038}, in which the dependence of the spin connection on the choice of frames on tetrahedrons can not be avoided in spatial scalar curvature naturally. 
 Second, note that the holonomy operator of the spin connection is constructed for the graph dual to an arbitrary cellular decomposition of the spatial manifold, but the continuum limit of the holonomy of the spin connection is checked only for the cubic graph. In fact, the cubic graph is special, since it is the one that gives the semiclassical consistent volume, as shown in \cite{Flori:2008nw}; moreover,  the continuum limit on cubic graph ensures a continuum field in a global coordinate system on the spatial manifold.  Nevertheless,  if we restrict our consideration to a local coordinate system  covering only the glued two faces dual to $e$,  as  in Ref. \cite{PhysRevD.87.024038}, the continuum limit of the holonomy of spin connection is also able to give the correct spin connection of continuum triad field for arbitrary non-cubic polyhedrons in Regge geometry.  
 Hence, by considering a proper regularization over the graph, the spatial scalar curvature operator $\hat{\tilde{R}}$ on the cubic graph can be also generalized to the graph dual to an arbitrary cellular decomposition, with the loop $\square_{e_I,e_J}$ in the operator $\widehat{h^{\Gamma}}_{\square_{e_I,e_J}}$ being reinterpreted as the minimal loop containing $e_I,e_J$ in the corresponding graph, and the volume operator being defined semi-classically consistent.}

 Our construction in this article suggests some further research directions. First, we can consider the semi-classical expansion of the expectation value of the spatial curvature operator on the gauge theory coherent states (GCS), as in Refs. \cite{Thiemann:2000bx}\cite{Giesel:2006um}. The main problem for this consideration is that we do not have a semiclassical expansion for the inverse volume operators currently.
Moreover, with Thiemann's trick there is no longer an inverse volume operator, but the spatial curvature operator still involves the inverse of the functions of flux operator. Thus, the application of the semi-classical expansion based on GCS and Algebraic quantum gravity (AQG) to the spatial curvature operator still needs further researches.
  Second, similar to the intrinsic curvature, the extrinsic curvature on a graph could also be expressed in terms of the twisted geometry variables \cite{Freidel:2010bw, PhysRevD.103.086016}.
 Then, noting that the twisted geometry provides a re-parametrization of the whole holonomy-flux phase space, one can further express the extrinsic curvature on a graph by the holonomy-flux variables. Compared to the previous regularization of the extrinsic curvature, which involves the commutator between the holonomy and the Euclidean term of the Hamiltonian constraint, this idea based on the twisted geometry may provide us with a simpler strategy to regularize the extrinsic curvature. Moreover, this idea may also be used to regularize the Lorentzian term $K^i_{[a}K^j_{b]}E^a_iE^b_j$ in the Hamiltonian constraint.
 Combining this new idea with the results of this article, one may introduce a new regularization scheme for the full Hamiltonian constraint based on the twisted geometry purely. Third,  the Hamiltonian constraint operator which contains this new spatial curvature operator as the Lorentzian term  can be used to derive standard LQC from full LQG as in Refs. \cite{Alesci:2013xd, Han:2019vpw, Kisielowski:2022wvk}, and derive the corresponding cosmological perturbation theory from full LQG as in Refs. \cite{Han:2020iwk, Han:2021cwb}. Besides we can also consider the study of implementation of the $\mu$-bar scheme for this new Lorentzian term, extending the results from \cite{ Alesci:2016rmn, Han:2019feb, Han:2021cwb,Giesel:2023euq}.

\section*{Acknowledgments}
This work is supported by the project funded by  the National Natural Science Foundation of China (NSFC) with Grants No. 12047519, No. 11875006 and No. 11961131013. HL is supported by the research grants
provided by the Blaumann Foundation. HL thanks the hospitality of the Beijing Normal University and Jinan University during his visit.

\appendix
\section{Proof of equation \eqref{gammax1ge}}\label{app:check_continuum}
Let us first separate the longitude and latitude components of $h_{e_1}^{\Gamma}$ as $h_{e_1}^{\Gamma}=\exp( \Gamma_{e_1}^{i}\tau_i)=\exp( \Gamma_{\text{lon}}^{i}\tau_i)\exp( \Gamma_{\text{lat}}^{i}\tau_i)$, where $\exp( \Gamma_{\text{lat}}^{i}\tau_i)$ rotates the vector $w^{v_1,\hat{x}^1}_{j}$ while $\exp( \Gamma_{\text{lon}}^{i}\tau_i)$ generates the rotation around the vector $w^{v_1,\hat{x}^1}_{j}$. Define the unit vectors $\tilde{w}_{v_1,\hat{x}_2j}:=\frac{w_{v_1,\hat{x}_2j}}{|w_{v_1,\hat{x}_2}|}$, $\tilde{w}_{v_1,\hat{x}_3j}:=\frac{w_{v_1,\hat{x}_3j}}{|w_{v_1,\hat{x}_3}|} $ and $\tilde{w}^{v_1,\hat{x}_1}_{j}:=\frac{w^{v_1,\hat{x}_1}_{j}}{|w^{v_1,\hat{x}_1}|}$ as in Fig.\ref{fig2}, where $|w_{v_1,\hat{x}_2}|$, $|w_{v_1,\hat{x}_3}|$ and  $|w^{v_1,\hat{x}_1}|$ are the module length of corresponding vectors. Then, one has
\begin{eqnarray}\label{proof1}
\exp( \Gamma_{e_1}^{i}\tau_i)&=&\exp( \Gamma_{\text{lon}}^{i}\tau_i)\exp( \Gamma_{\text{lat}}^{i}\tau_i) \\\nonumber
&=&(\mathbb{I}+\theta_{\hat{x}_2}\tilde{w}_{v_1,\hat{x}_2j}\tau^j +\theta_{\hat{x}_3}\tilde{w}_{v_1,\hat{x}_3j}\tau^j+\mathcal{O}(\mu^2) )(\mathbb{I} +\theta_{\hat{x}_1}\tilde{w}^{v_1,\hat{x}_1}_{j}\tau^j+\mathcal{O}(\mu^2))
\end{eqnarray}
for small $\mu = \max\{\theta_{\hat{x}_i}\} \ll 1 $, where $\theta_{\hat{x}_2}\tilde{w}_{v_1,\hat{x}_2j}\tau^j $, $\theta_{\hat{x}_3}\tilde{w}_{v_1,\hat{x}_3j}\tau^j $ and $\theta_{\hat{x}^1}\tilde{w}^{v_1,\hat{x}_1}_{j}\tau^j$ are the infinite small generators of the rotations around the vectors  $\tilde{w}_{v_1,\hat{x}_2j}$, $\tilde{w}_{v_1,\hat{x}_3j}$ and $\tilde{w}^{v_1,\hat{x}_1}_{j}$.
Let us first adapt the rotation of the polyhedra dual to $v_2$ in Fig.\ref{fig2} to the generators $\theta_{\hat{x}_2}\tilde{w}_{v_1,\hat{x}_2j}\tau^j $ at $h_{e_1}^{\Gamma}=\mathbb{I}$, one has
\begin{eqnarray}\label{proof2}
\theta_{\hat{x}_2}\tilde{w}_{v_1,\hat{x}_2}^{i}
&=&\epsilon^{ijk}w_k^{v_1,\hat{x}_1}(w^{v_1,\hat{x}_3}_jw_{v_1,\hat{x}_1l} (w^l_{v_2,\hat{x}_3}-w^l_{v_1,\hat{x}_3}))+\mathcal{O}(\mu^2),
\end{eqnarray}
where we use $w_{v_2,\hat{x}_3j} -w_{v_1,\hat{x}_3j} \simeq \theta_{\hat{x}_2} \epsilon_{jkl} \tilde{w}_{v_1,\hat{x}_2}^{k} w_{v_1,\hat{x}_3}^l$ for the generators $\theta_{\hat{x}_2}\tilde{w}_{v_1,\hat{x}_2j}\tau^j$.  Similarly,  by adapting the rotation of the polyhedra dual to $v_2$ in Fig.\ref{fig2} to the generators $\theta_{\hat{x}_3}\tilde{w}_{v_1,\hat{x}_3j}\tau^j $ and $\theta_{\hat{x}^1}\tilde{w}^{v_1,\hat{x}_1}_{j}\tau^j$ at $h_{e_1}^{\Gamma}=\mathbb{I}$ respectively, one can get
\begin{eqnarray}\label{proof3}
\theta_{\hat{x}_3}\tilde{w}_{v_1,\hat{x}_3}^i
&=&\epsilon^{ijk}w_k^{v_1,\hat{x}_1}(w^{v_1,\hat{x}_2}_jw_{v_1,\hat{x}_1l}(w^l_{v_2,\hat{x}_2} -w^l_{v_1,\hat{x}_2})) +\mathcal{O}(\mu^2)
\end{eqnarray}
with  $w_{v_2,\hat{x}_2j} -w_{v_1,\hat{x}_2j} \simeq \theta_{\hat{x}_3} \epsilon_{jkl} \tilde{w}_{v_1,\hat{x}_3}^{k} w_{v_1,\hat{x}_2}^l$ for the generators $\theta_{\hat{x}_3}\tilde{w}_{v_1,\hat{x}_3j}\tau^j $,
and
\begin{eqnarray}\label{proof4}
\theta_{\hat{x}_1}\tilde{w}^{v_1,\hat{x}_1i}&=&-\epsilon^{ijk}
w_k^{v_1,\hat{x}_3}(w_{v_2,\hat{x}_3j} -w_{v_1,\hat{x}_3j})
\nonumber\\
&&+\epsilon^{ijk}w_k^{v_1,\hat{x}_1}(w^{v_1,\hat{x}_2}_jw_{v_1,\hat{x}_1l}(w^l_{v_2,\hat{x}_2} -w^l_{v_1,\hat{x}_2})) +\mathcal{O}(\mu^2)\nonumber\\
&=&-\epsilon^{ijk}w_k^{v_1,\hat{x}_2}(w_{v_2,\hat{x}_2j}-w_{v_1,\hat{x}_2j}) \\
 &&+\epsilon^{ijk}w_k^{v_1,\hat{x}_1}(w^{v_1,\hat{x}_3}_jw_{v_1,\hat{x}_1l} (w^l_{v_2,\hat{x}_3}-w^l_{v_1,\hat{x}_3})) +\mathcal{O}(\mu^2) \nonumber
\end{eqnarray}
with $w_{v_2,\hat{x}_3}^l -w_{v_1,\hat{x}_3}^l \simeq \theta_{\hat{x}^1}\epsilon^{jkl}\tilde{w}^{v_1,\hat{x}_1}_{j}w_{v_1,\hat{x}_3k} $ and $w^l_{v_2,\hat{x}_2} -w^l_{v_1,\hat{x}_2} \simeq \theta_{\hat{x}^1}\epsilon^{jkl}\tilde{w}^{v_1,\hat{x}_1}_{j}w_{v_1,\hat{x}_2k}$  for the generator $\theta_{\hat{x}^1}\tilde{w}^{v_1,\hat{x}_1}_{j}\tau^j$.
Finally, combine the results of Eqs.\eqref{proof1}, \eqref{proof2}, \eqref{proof3} and \eqref{proof4}, the Eq.\eqref{gammax1ge} can be verified immediately.

\bibliographystyle{unsrt}

\bibliography{ref}


\end{document}